\documentclass[preprint,prc,showpacs,preprintnumbers,amsmath,amssymb,floatfix]{revtex4}
\usepackage{amsmath}
\usepackage{amssymb}
\usepackage{graphicx}
\usepackage{dcolumn}
\usepackage{bm}
\usepackage{ulem} 
\usepackage[usenames]{color}
\usepackage{epstopdf}
\usepackage{epsfig}
\usepackage{float}
\usepackage{subfigure}
\usepackage{multirow}
\usepackage[version=3]{mhchem}
\usepackage{hyperref}
\usepackage{array}

\newcommand{\nc}{\newcommand}       
\newcommand{\renc}{\renewcommand}   
\nc{\vc}[1] {\mbox{\boldmath $#1$}} 
\nc{\del}       {\partial}              
\nc{\bra}       {\langle}               
\nc{\ket}       {\rangle}               
\nc{\bras}[1]   {\langle #1|}           
\nc{\kets}[1]   {|#1\rangle}            
\nc{\mapleft}[1]{           
	\smash{\mathop{\,          %
			\hbox to 1.5cm{\rightarrowfill}\, }\limits_{#1}}}
\nc{\beq}     {\begin{eqnarray}} \nc{\eeq}    {\end{eqnarray}}
\nc{\nn}      {\\\nonumber} \nc{\vs}      {\vspace{-0.275cm}}
\nc{\fra}    {\frac{1}{2}}
\nc{\mb}        {\mathbf}
\nc{\wt} {\widetilde}
\renc\thesubsection{\arabic{subsection}}

\begin{document}
	\title{Properties of nuclear matter in relativistic Brueckner-Hartree-Fock model with high-precision charge-dependent potentials}
	\author{Chencan Wang}
	\affiliation{School of Physics, Nankai University, Tianjin 300071,  China}
	\author{Jinniu Hu}
	\email{hujinniu@nankai.edu.cn}
	\affiliation{School of Physics, Nankai University, Tianjin 300071,  China\\
	Strangeness Nuclear Physics Laboratory, RIKEN Nishina Center, Wako, 351-0198, Japan}
	\author{Ying Zhang}
	\affiliation{Department of Physics, Faculty of Science, Tianjin University, Tianjin 300072, China\\
	Strangeness Nuclear Physics Laboratory, RIKEN Nishina Center, Wako, 351-0198, Japan}
	\author{Hong Shen}
	\affiliation{School of Physics, Nankai University, Tianjin 300071,  China}
	\date{\today}
	
	\begin{abstract}
Properties of nuclear matter are investigated in the framework of relativistic Brueckner-Hartree-Fock model with the latest high-precision charge-dependent Bonn (pvCD-Bonn) potentials, where the coupling between pion and nucleon is adopted as pseudovector form. These realistic pvCD-Bonn potentials are renormalized to effective nucleon-nucleon ($NN$) interactions, $G$ matrices. They are obtained by solving the Blankenbecler-Sugar (BbS) equation in nuclear medium. Then, the saturation properties of symmetric nuclear matter are calculated with pvCD-Bonn A, B, C potentials. The energies per nucleon are around $-10.72$ MeV to $-16.83$ MeV at saturation densities, $0.139$ fm$^{-3}$ to $0.192$ fm$^{-3}$ with these three potentials, respectively. It clearly demonstrates that the pseudovector coupling between pion and nucleon can generate reasonable saturation properties comparing with pseudoscalar coupling. Furthermore, these saturation properties have strong correlations with the tensor components of $NN$ potentials, i.e., the $D$-state probabilities of deuteron, $P_D$ to form a relativistic Coester band. The equations of state of pure neutron matter from pvCD-Bonn potentials are almost identical, since the prominent difference of pvCD Bonn potentials are the components of tensor force, which provides very weak contributions in the case of total isospin $T=1$. In addition, the charge symmetry breaking (CSB) and charge independence breaking (CIB) effects are also discussed in nuclear matter from the partial wave contributions with these high-precision charge-dependent potentials. In general, the magnitudes of CSB from the differences between $nn$ and $pp$ potentials are about $0.05$ MeV, while those of CIB are around $0.35$ MeV from the differences between $np$ and $pp$ potentials. Finally, the equations of state of asymmetric nuclear matter are also calculated with different asymmetry parameters. It is found that the effective neutron mass is larger than the proton one in neutron-rich matter.
	\end{abstract}

\pacs{21.10.Dr,  21.60.Jz,  21.80.+a}

\maketitle

\section{Introduction}
The infinite nuclear matter is a fundamental study subject in nuclear physics, where protons and neutrons compose a uniform many-body system in the nuclear matter. Due to the translation invariance and rotation invariance, their wave functions are regarded as plane waves~\cite{baldo99}. Although the nuclear matter is a hypothetical substance, the saturation properties of symmetric nuclear matter can be extracted from experimental observations in the central region of heavy nuclei~\cite{klupfel09,rocamaza18}. Furthermore, the equation of state of neutron-rich matter plays very important roles in the investigations of many astrophysical processes, such as, supernova explosion, neutron star cooling, binary neutron star merger, and so on~\cite{oertel17,abbott17a,abbott17b,abbott17c,goldstein17,abbott18a}. Recently, with the worldwide development of radioactive facilities, many neutron-rich nuclei were discovered, where the isospin properties of nuclear matter are hopefully extracted, i.e., the symmetry energy and its slope~\cite{li08,li19}. Moreover, the central density of compact star is closed to five times of nuclear saturation density, which is far beyond the present experimental abilities. Therefore, the properties of nuclear matter from theoretical researches are highly demanded from both the investigations of neutron-rich nuclei and nuclear astrophysics~\cite{shen98a,shen98b,shen02,shen11}.  

Due to the complexity of nucleon-nucleon ($NN$) potential, the study of nuclear matter is not as straightforward as the electron gas in condensed matter physics, although both of them are considered as uniform systems. The first calculation on the properties of nuclear matter was achieved by Euler eighty years ago with second-order perturbation theory based on Hartree-Fock approximation, where the $NN$ potential was chosen as a Gaussian function~\cite{euler37}. With abundant experimental data of $NN$ scattering since 1940s, Jastrow proposed that there was a very strong repulsive core at short-range distance between two nucleons in free space~\cite{jastrow51}. It means that the nuclear many-body system cannot be treated in the viewpoint of perturbation theory with the $NN$ interaction derived from the $NN$ scattering data, i.e., realistic $NN$ potential. Therefore, various nuclear many-body methods were developed to study the nuclear matter in the past seventy years. 

The strong repulsion of $NN$ potential at the short-range distance must be renormalized in nuclear medium to generate the bound states of finite nuclei and saturation properties of symmetric nuclear matter. The earliest renormalization method was proposed by Brueckner {\it et al.}, where the repulsion can be removed by summations of all ladder diagrams included in the nuclear medium $NN$ scattering process~\cite{brueckner54,bethe56}. The realistic $NN$ interaction will be replaced by a density-dependent potential, $G$ matrix. It can be used to describe the nuclear many-body system in Hartree-Fock approximation reasonably. Meanwhile, a variational method was shown by Jastrow through considering correlation functions to transfer the trial wave functions to the exact ones~\cite{jastrow55}.

With the rapid developments of high-precision $NN$ potentials and the computational techniques, many advanced nuclear many-body methods with realistic $NN$ potentials were developed in nonrelativistic framework, such as Brueckner-Hartree-Fock method~\cite{li06,baldo07,baldo16}, quantum Monte Carlo methods~\cite{akmal98,carlson15}, self-consistent Green's function method~\cite{dickhoff04}, coupled-cluster method~\cite{hagen14a,hagen14}, many-body perturbation theory~\cite{carbone13,carbone14,drischler14}, functional renormalization group (FRG) method~\cite{drews15,drews16}, lowest order constrained variational method~\cite{modarres93}, and so on. These methods can more or less obtain the saturation behaviors of symmetric nuclear matter with modern high-precision $NN$ potentials, like Reid93 potential, Nijmegen potential~\cite{stoks94}, AV18 potential~\cite{wiringa95}, CD-Bonn potential~\cite{machleidt01}, chiral N$^3$LO potentials~\cite{entem03,epelbaum05} and chiral N$^4$LO potentials~\cite{epelbaum15a,epelbaum15b,entem15,entem17,reinert18}. However, all saturation properties from these calculations cannot reproduce the empirical data, $E/A=-16\pm1$ MeV at $n_0=0.16\pm0.01$ fm$^{-3}$ only with two-body nuclear force. In order to reproduce the reasonable saturation properties, the three-body nucleon force must be introduced in these nonrelativistic frameworks to provide additional repulsion contributions~\cite{li06,hu17,sammarruca18,logoteta19}.

In 1980s, the relativistic version of Brueckner-Hartree-Fock method was firstly proposed by Anastasio {\it et al.}~\cite{ansatasio83}, then developed by Horowitz {\it et al.}~\cite{horowitz87} and Brockmann {\it et al.}~\cite{brockmann90} In the relativistic Brueckner-Hartree-Fock (RBHF) model, a repulsive contribution is obtained from the relativistic effect, which can properly describe the nuclear saturation properties with two-body realistic $NN$ potential. 
Li {\it et al.} also verified that the contributions from three-body force and $Z$ diagram, i.e. the nucleon-antinucleon excitation from the relativistic effect are partially in accord with each other~\cite{lizh08}, since the nucleon-antinucleon excitation affects the energy of nuclear matter in RBHF model via the second-order term of scalar meson, which can be regarded as one component of the microscopic three-body force generated by the two-meson exchange between nucleon excitation states. Furthermore, the RBHF model was also applied to investigate the superfluity of nuclear matter, properties of neutron star and help to fit the free parameters of nuclear density functional theories~\cite{alonso03,krastev06,sammarruca10,dalen10}. Recently, Shen {\it et al.} realized a fully self-consistent calculation of RBHF model in finite nuclei system and extended this framework on the neutron drops~\cite{shen16,shen17,shen19}. The exact treatment for the angular integration of the center-of-mass momentum in asymmetric nuclear matter was also worked out by Tong {\it et al.}~\cite{tong18} within RBHF model.

In RBHF model, the nuclear medium effect must be taken into account in the $NN$ potential. Therefore, only few $NN$ interactions can be adopted, such as Bonn potentials~\cite{machleidt87,machleidt89}. With a large number of two-nucleon scattering data, many high-precision $NN$ potentials were proposed based on the charge-dependent partial wave analysis from 1990s, as mentioned before, like Reid93~\cite{stoks94}, AV18~\cite{wiringa95}, CD-Bonn potentials~\cite{machleidt01}, and so on. In addition, the chiral $NN$ potentials derived from the chiral perturbation theory were also developed rapidly. The high-precision chiral $NN$ potentials, N$^4$LO potentials~\cite{epelbaum15a,epelbaum15b,entem15,entem17,reinert18}, were already presented up to the fifth chiral expansion order. {These state-of-the-art chiral potentials have been widely applied to describe the structures of finite nuclei and the saturation properties of infinite nuclear matter. When the three-body and four-body forces obtained from chiral perturbation theory systematically are included, the properties of light nuclei and nuclear matter were reproduced perfectly below the breakdown scales~\cite{epelbaum15b,hu17,sammarruca18,logoteta19}. In such controlled hierarchy, the uncertainties from the few-body forces can be nicely estimated. We discussed such truncation errors and breakdown scale with Bayesian method for symmetric nuclear matter and pure neutron matter with latest chiral potentials~\cite{hu19}. It was found that the breakdown scale of these chiral potentials is around $600$ MeV and the uncertainties from high-order potentials increase with density. With such investigation, the properties of nuclear matter below $0.4$ fm$^{-3}$ should be believable for present chiral potentials. However, the study of compact star requires the equation of state of nuclear matter above  $0.8$ fm$^{-3}$ . Therefore, it is very import to adopt an available many-body method and high-precision $NN$ potentials for a better description of the nuclear matter especially in the high density region.}

In principle, the high-precision CD-Bonn potential with the same framework of Bonn potential can be used in RBHF model. However, its pseudoscalar coupling scheme between pion and nucleon in relativistic framework will generate a very strong attractive contribution and thus can not reproduce the empirical saturation properties. Therefore, we attempted to use pseudovector coupling instead of the pseudoscalar one between pion and nucleon. New pvCD-Bonn potentials were obtained by fitting the $NN$ phase shifts from the Nijmegen partial wave analysis, which can be used in the RBHF model~\cite{wang19}. 

In this work, properties of nuclear matter will be calculated in RBHF model with the latest pvCD-Bonn potentials. The exact angular integration of center-of-mass momentum will also be achieved. There are three pvCD-Bonn potentials with different tensor components, whose effects to saturation properties will be investigated. Furthermore, the charge symmetry breaking (CSB) and charge independent breaking (CIB) effects also will be discussed, which were calculated by Sammarruca {\it et al.} in Brueckner-Hartree-Fock method with the original CD-Bonn potential~\cite{sammarruca12}. This paper is arranged as follows: in section~\ref{Theory}, the necessary formulas of RBHF model for asymmetric nuclear matter will be introduced. In section~\ref{Results}, properties of nuclear matter calculated by the new pvCD-Bonn potentials will be presented, including the equations of state of nuclear matter, the single-particle potentials, partial-wave contributions to the potential energy, CSB and CIB effects in nuclear matter, and so on. In section~\ref{End},  summaries and conclusions will be shown. The supplement derivations of in-medium Blankenbecler-Sugar (BbS) equation and the numerical details involved in practical calculations are given in the appendices.

\section{The relativistic Brueckner-Hartree-Fock model in nuclear matter}\label{Theory}
In RBHF model, the single-nucleon energy in nuclear matter, $E_\tau$ is given by a Dirac equation with a single-particle potential $U_\tau$~\cite{brockmann90,tong18},
         \begin{equation}\label{DiracEq}
           (\bm{\alpha}\cdot \mb{p}  + \beta M_\tau + \beta U_\tau ) u_\tau(\mathbf{p},s)
          = E_\tau(\mb{p})  u_\tau(\mathbf{p},s),
        \end{equation}
where the subscript $\tau=p~,n$ indicates proton or neutron. $M_\tau$ is the nucleon mass and $u_\tau(\mathbf{p},s)$ is the spinor solution of this Dirac equation with momentum $\mathbf{p}$ and spin $s$. The single-particle potential in nuclear matter can be expressed as,
        \begin{equation} \label{SP-Pot-Full}
            U_\tau= U_{\tau,\text{s}} +\gamma^0  U^0_{\tau,\text{v}}  -\bm{\gamma}\cdot
            \mb{p} U^i_{\tau, \text{v}},
        \end{equation}
which is decomposed into a scalar component $U_{\tau,\text{s}}$ and a vector one $U_{\tau,\text{v}}$ due to the translational invariance and rotation invariance of infinite nuclear matter. The available investigations showed that the momentum dependence of scalar and vector potentials are very weak. Furthermore, the magnitude of the spacelike component of vector potential, $U^i_{\tau, \text{v}}$, is negligible comparing to the timelike one, $U_{\tau, \text{v}}^0$, and the scalar potential, $U_{\tau,\text{s}}$~\cite{sammarruca10}. 

Here, it must be emphasized that actually there are two schemes in RBHF model to determine the Dirac structure of the nucleon self-energy. The first one is what we have done following the framework of  Brockmann and Machleidt~\cite{brockmann90}, where the momentum dependence of the Dirac components of self-energy are neglected and the components are derived from the momentum dependence of the single-particle energies. The second one is evaluating the Dirac structure of the nucleon self-energy using a projection technique and keep the momentum dependence~\cite{dalen10}. These two schemes can yield rather similar results for scalar and vector potentials in symmetric nuclear matter. However, as an example we mention that the isospin-dependent behavior of the effective nucleon masses in asymmetric nuclear matter, related to their scalar potentials, are completely opposite~\cite{dalen10a} . Therefore, in this work, we will use the Brockmann-Machleidt scheme to discuss the properties of nuclear matter with pvCD-Bonn potentials and investigate them in the future using the project method.

Therefore, the single-particle potential is approximately written as
       \begin{equation} \label{SP-Pot-Approx}
            U_\tau \approx U_{\tau,\text{s}} +\gamma^0  U_{\tau,\text{v}}.
        \end{equation}
 Within such approximation, the Dirac equation \eqref{DiracEq} in nuclear medium will be reduced as, 
        \begin{equation} \label{DiracEq-Inmedium}
            (\bm{\alpha}\cdot \mb{p}  + \beta M^*_\tau) u_\tau(\mathbf{p},s)
            = E^*_\tau(\mb{p})  u_\tau(\mathbf{p},s)
        \end{equation}
with effective nucleon mass and energy dressed in nuclear medium,
        \begin{equation} \label{Eff-ME}
            M^*_\tau = M_\tau + U_{\tau,\text{s}}, \quad
            E^*_\tau = E - U_{\tau,\text{v}}.
        \end{equation}
The wave function of Dirac equation in nuclear matter~\eqref{DiracEq-Inmedium} can be solved analytically as a plane wave,
        \begin{equation} \label{Spinor-Covar}
            u_\tau(\mathbf{p},s)= \sqrt{\frac{E^*_\tau+M_\tau^*}{2M^*_\tau}}
            \left(\begin{array}{c}
            1 \\
            \frac{\bm{\sigma}\cdot\mb{p}}{M^*_\tau+E^*_\tau} \\
            \end{array}\right) \chi_s,
        \end{equation}
where $\chi_s$ stands the spin wavefunction for $s$ state and $E_\tau^*(\mb{p})=\sqrt{\mb{p}^2 + M_\tau^{*2}}$ is the in-medium on-shell single-particle energy. The normalization condition of spinor is $\bar{u}(\mb{p},s)u(\mb{p},s) =1$, here.

 The nucleon state vector can be expressed as $|\mb{p},s\rangle = u(\mb{p},s)$ and with its conjugated vector $\langle \mb{p},s | = u^\dag(\mb{p},s)$. Hence, there will be an extra factor $M^*/E^*$ to normalize the nucleon state due to the choice of $\bar{u}(\mb{p},s)u(\mb{p},s) =1$,
        \begin{equation}\label{Norm}
            \frac{M^*}{E^*} \langle \mb{p},s |\mb{p},s\rangle=1.
        \end{equation}
 The expectation value of single-particle potential can be evaluated within nucleon state vectors,
        \begin{equation} \label{SP-Pot}
            U_\tau(p) =  \frac{M^*_\tau}{E_\tau^*}\langle \mb{p},s|\beta U_\tau|\mb{p},s\rangle_\tau
            = \frac{M_\tau^*}{E_\tau^*} U_{\tau,\text{s}} + U_{\tau,\text{v}},
        \end{equation}
 where the $U_{\tau,\text{s}}$ and $U_{\tau,\text{v}}$ are regarded as momentum independent. Their detailed values should be determined by the $NN$ interaction.    

In RBHF model, the realistic $NN$ interactions are replaced by effective $G$ matrices due to the nuclear medium effect, where the strong repulsion of realistic $NN$ potential at short-range distance is renormalized through summations of two-nucleon scattering ladder diagrams. These diagrams can be contracted as an integral equation in free space, i.e., Bethe-Salpeter equation~\cite{salpeter51} in four-dimension space. It is usually reduced to a three-dimension equation to simplify the calculations. There are many reduction schemes, such as, Blankenbecler-Sugar (BbS) equation~\cite{blankenbecler66}, Thompson equation~\cite{thompson70}, Kadyshevsky equation~\cite{kadyshevsky68}, and so on. Since the pvCD-Bonn potentials were obtained in the framework of BbS equation at free space~\cite{wang19}, the $G$ matrix in present RBHF model should be the solutions of BbS equation in nuclear medium derived at appendix A, which is written as,
        \begin{equation}\label{BbS-Gmatrix}
            G_{\tau_1\tau_2}(\mathbf{q}',\mathbf{q}|\mathbf{P}) = V_{\tau_1\tau_2}
            (\mathbf{q}',\mathbf{q}) + \int \frac{d^3k}{(2\pi)^3}V_{\tau_1\tau_2}
            (\mathbf{q}',\mathbf{k})\frac{2W_k}{W_0+W_k}
            \frac{Q_{\tau_1\tau_2}(\mathbf{k},\mathbf{P})}{W_0 - W_k}
            G_{\tau_1\tau_2}(\mathbf{k},\mathbf{q}|\mathbf{P}),
        \end{equation}
where $\mathbf{q}$ , $\mathbf{k}$,  and $\mathbf{q}'$  are initial, intermediate, and final relative momenta, respectively. $\mathbf{P}$ is the two-nucleon center-of-mass momentum in nuclear matter rest frame. $\tau$ denotes the third component of nucleon isospin. The transformations between nuclear matter rest frame and center-of-mass frame are
        \begin{equation} \label{Rest2CM}
            \mb{k} = \frac{\mb{p}'-\mb{p}}{2}, \qquad \mb{P} = \frac{\mb{p}+\mb{p}'}{2}.
        \end{equation}
In BbS equation~\eqref{BbS-Gmatrix}, $V_{\tau_1\tau_2}$ and $G_{\tau_1\tau_2}$ are related to the covariant amplitudes $\bar{V}_{\tau_1\tau_2}$ and $\bar{G}_{\tau_1\tau_2}$ with additional factors derived from the normalization convention in Eq.~\eqref{Norm}, which are expressed as
        \begin{equation}\label{Gbar2G}
        V_{\tau_1\tau_2}
        =\frac{M^*_\tau}{E^*_{\tau_1}}\bar{V}_{\tau_1\tau_2}\frac{M^*_{\tau_1}}{E^*_{\tau_2}},
        \quad
        G_{\tau_1\tau_2}
        =\frac{M^*_\tau}{E^*_{\tau_1}}\bar{G}_{\tau_1\tau_2}\frac{M^*_{\tau_1}}{E^*_{\tau_2}}.
        \end{equation}

To prevent the scattering states into the Fermi sea, a Pauli blocking operator
        \begin{equation}\label{PauliBlockOperator}
            Q_{\tau_1\tau_2}(\mathbf{k},\mathbf{P}) = \left\{
            \begin{array}{cl}
            1  &   \quad  (|\mathbf{P}+\mathbf{k}|>k_F^{\tau_1}~
            \text{and}~|\mathbf{P}-\mathbf{k}|>k_F^{\tau_2}), \\
            0  &   \quad  (\text{otherwise}),
            \end{array}
            \right.
        \end{equation}
is taken into account comparing to the free BbS equation, where $k_F^{\tau}$ represents the Fermi momentum for nucleon $\tau$.
Furthermore, $W_0 = E^*_{\tau_1}(\mathbf{P}+\mathbf{q}) + E^*_{\tau_2}(\mathbf{P}-\mathbf{q})$ and $W_k = E^*_{\tau_1}(\mathbf{P}+\mathbf{k}) + E^*_{\tau_2}(\mathbf{P}-\mathbf{k})$ are the starting and intermediate energies respectively.

The equation of state (EOS) of nuclear matter is a function of  baryon number density, $n_b$ and asymmetry parameter, $\alpha$, 
        \begin{equation}\label{NM-Params}
            n_b = n_p + n_n , \qquad \alpha =\frac{n_n-n_p}{n_b},
        \end{equation}
where $n_p,~n_n$ are the baryon densities of proton and neutron. The averaged Fermi momentum is defined as $k_F = (3\pi^2 n_b/2)^\frac{1}{3}$, therefore Fermi momenta of proton and neutron are shown as
        \begin{equation}\label{}
        k_F^p = (1-\alpha)^\frac{1}{3}k_F, \quad \text{and}\quad 
        k_F^n = (1+\alpha)^\frac{1}{3}k_F.
        \end{equation}

When the Hartree-Fock approximation is applied, the single-particle potential of nucleon with isospin $\tau$ is evaluated through $G$ matrix
        \begin{equation}
        \label{HF-SP-Pot}
        \begin{aligned}
            U_\tau(p) &= \sum_{\tau'}\sum_{ss'}\int^{p' \leqslant k^\tau_F}\frac{d^3p'}{(2\pi)^3}
            \langle \mb{p}s,\mb{p}'s' |G_{\tau\tau'}| \mb{p}s,\mb{p}'s'
            -\mb{p}'s',\mb{p}s\rangle.
        \end{aligned}
        \end{equation}
In asymmetric nuclear matter, when charge symmetry breaking (CSB) and charge independence breaking (CIB) effects are considered, the $G$ matrices are divided by $G_{pp},~G_{np},~G_{pn}$, and $G_{nn}$. The corresponding single-particle potential for proton or neutron can be written as 
        \begin{equation}\label{Up-Un}
            U_\tau = U_{\tau p}  + U_{\tau n}. 
        \end{equation}
        
At a given density, a self-consistent numerical calculation for singe-particle potential via Eq.~\eqref{HF-SP-Pot} is started with initial scalar and vector potentials for proton and neutron. The $G$ matrices are solved with two-body $NN$ potential including the nuclear medium effect from Eq.~\eqref{BbS-Gmatrix}. Then the new scalar and vector potentials can be extracted through Eq.~\eqref{SP-Pot}. With the new scalar potentials, next-round calculation is worked out, until proton and neutron scalar potentials are converged at an acceptable accuracy. Finally, the energy per nucleon of nuclear matter at a fixed $n_b$ and $\alpha$ is evaluated by
        \begin{equation}\label{E/A}
        \begin{aligned}
            \frac{E}{A} &= \frac{1}{n_b}
            \sum_{\tau,s}\int^{p\leqslant k_F^\tau}\frac{d^3 p}{(2\pi)^3}\frac{M^*_\tau}{E^*_\tau}
            \langle \mb{p},s|\bm{\alpha}\cdot \mb{p} + \beta M_\tau |\mb{p},s\rangle -
            \frac{1-\alpha}{2}M_p - \frac{1+\alpha}{2}M_n\\
            & + \frac{1}{2 n_b} \sum_{\tau\tau'}\sum_{ss'} \int^{p\leqslant k_F^\tau}
            \frac{d^3p}{(2\pi)^3}\int^{p'\leqslant k_F^{\tau'}} \frac{d^3p'}{(2\pi)^3}
            \langle\mb{p}s,\mb{p}'s'|G_{\tau\tau'}|\mb{p}s,\mb{p}'s'-\mb{p}'s',\mb{p}s\rangle.
        \end{aligned}
        \end{equation}
In practical calculations, the variables, $\mb{p}$ and $\mb{p}'$, in integrals~\eqref{HF-SP-Pot} and \eqref{E/A} are replaced by $\mb{q}$ and $\mb{P}$. With further partial wave decomposition, $G$ matrices are projected into $LSJ$ representation. The solid angle dependence is removed in these integrals. The detailed expressions for numerical calculations are explicitly given in Appendix B.

 \section{Results and discussions}\label{Results}
 \subsection{The properties of symmetric nuclear matter and pure neutron matter}
   In our previous work~\cite{wang19},  three charge-dependent Bonn potentials, named as pvCD-Bonn A, B, C, with pseudovector (PV) coupling between nucleon and pion were obtained by fitting the $NN$ scattering phase shifts driven from Nijmegen partial wave analysis. These three potentials are almost identical except their tensor components due to the different $\pi N$ coupling strengths. The $D$-state probabilities of deuteron, $P_D$ generated by pvCD-Bonn A, B, C potentials are $4.22\%,~5.45\%$, and $6.05\%$, respectively.
                \begin{figure}[http]
                 \centering
                    \includegraphics[scale=0.6]{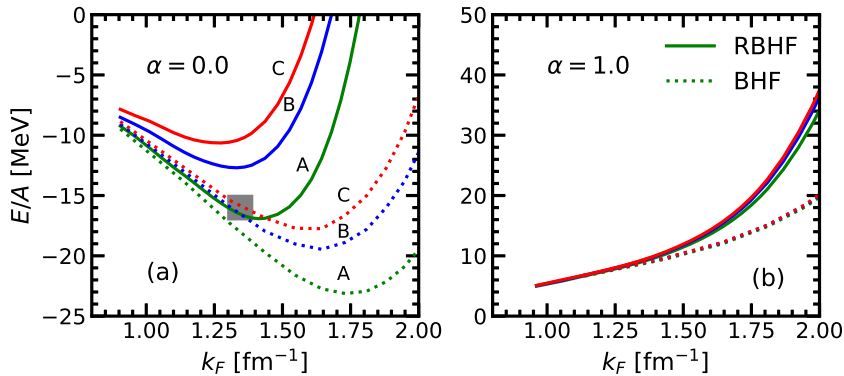}\\
                    \caption{Equations of state of symmetric nuclear matter and pure neutron matter calculated by BHF and RBHF models with pvCD-Bonn A, B, C potentials. The panel~(a) for symmetric nuclear matter with the rectangular patch labeling  the empirical saturation region. Panel~(b) for pure neutron matter.}\label{BHFvsDBHF}
                \end{figure}

In panels (a) and (b) of Fig.~\ref{BHFvsDBHF}, the energies per nucleon, $E/A$ as functions of Fermi momentum, $k_F$, i.e., equations of state, for symmetric nuclear matter and pure neutron matter, calculated by RBHF model are plotted within pvCD-Bonn potentials as solid curves, respectively. It can be found that saturation properties of symmetric nuclear matter from pvCD-Bonn A are closest to the empirical values shown as the rectangular area among three potentials. Its energy per nucleon, $-16.83$ MeV satisfies the value extracted from the mass formula of finite nuclei, while the corresponding saturation density, $n_0$ is $0.192$ fm$^{-3}$, that is higher than the normal one, $0.16\pm0.01$ fm$^{-3}$. On the other hand, the saturation density from pvCD-Bonn B potential, $n_0=0.158$ fm$^{-3}$, however its energy per nucleon at saturation density is just $-12.91$ MeV. These differences between calculations of RBHF model and empirical values may be caused by that the non-nucleon degree of freedom, like the $\Delta$-isobar should be included in the $NN$ interaction as shown in recent works~\cite{deltuva03,logoteta16,ekstrom18}. The pvCD-Bonn A potential owns the weakest tensor component in three potentials and generates the largest saturation binding energy. On the whole, these results and conclusions are similar with those from Bonn potentials by Brockmann and Machleidt~\cite{brockmann90}.  For the pure neutron matter, the differences of energy per nucleon among three potentials are quite small. It is because that the tensor effect is very weak in pure neutron matter~\cite{hu10,wang12,zhang19}, where the total isospin of two neutron is $T=1$ and contribution of tensor force is largely suppressed.

Furthermore, the equations of state of symmetric nuclear matter and pure neutron matter are also obtained in the nonrelativistic framework of BHF model with the same $NN$ potentials, which are given as dashed curves in Fig.~\ref{BHFvsDBHF}. At low density region, their energies per nucleon are very similar with those in RBHF model. With the density increasing, the relativistic effect from the nucleon-antinucleon excitation, becomes obvious, provides strong repulsive contributions, and leads to reasonable saturation properties of symmetric nuclear matter, which plays a similar role with the three-body force in the nonrelativistic {\it ab initio} approaches. Actually, Li {\it et al.} also confirmed that the three-body force and nucleon-antinucleon $Z$-diagram create the equivalent contributions in nuclear matter~\cite{lizh08}. Furthermore, the free nucleon mass in $NN$ potential will be replaced by an effective nucleon mass, derived from the scalar potential to achieve the self-consistent RBHF calculation.               
                              
                \begin{figure}[http]
                    \centering
                    \includegraphics[scale=0.6]{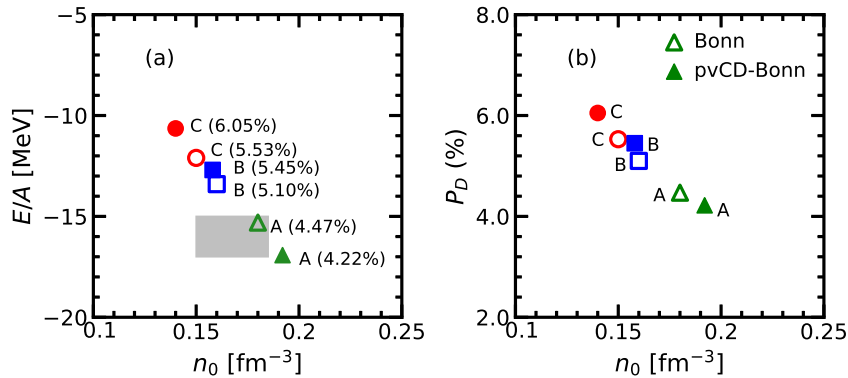}\\
                    \caption{A relativistic Coester band. The rectangle patch indicates empirical
                    saturation region. The open patterns refer to the saturation properties from Bonn potentials, while the solid ones correspond to those from pvCD-Bonn potentials. }
                    \label{CoesterBand}
                \end{figure}
            
In available investigations, the saturation properties of symmetric nuclear matter from BHF model included strong correlations with the strengths of tensor force in realistic two-body $NN$ interactions, which can be represented by the $D$-state probability of deuteron, $P_D$. In general, the weaker tensor strength (smaller $P_D$) generates a larger saturation density and deeper binding energy. This correlation relation was so-called "Coester band"~\cite{coester70}. In the panel (a) of Fig.~\ref{CoesterBand}, the saturation densities and the corresponding energies per nucleon of symmetric nuclear matter from pvCD-Bonn A, B, C potentials and Bonn A, B, C potentials in RBHF model are shown. There is a fine linear relationship between them with different $P_D$ and this relativistic Coester band can cross over the empirical saturation region. Generally speaking, the $NN$ potential with lower $P_D$, around $4\%-5\%$ is preferred to generate relatively reasonable saturation properties. The correlation between $D$-state probability and saturation density is shown in the panel (b) with different potentials. A larger tensor component results in a smaller saturation density. This is because that the tensor force provides the largely attractive contributions in low density region for symmetric nuclear matter and makes the saturation density go back, while the short-range correlation becomes more important with density increasing~\cite{hu13}.  

\begin{figure}[http]
	\centering
	\includegraphics[scale=0.6]{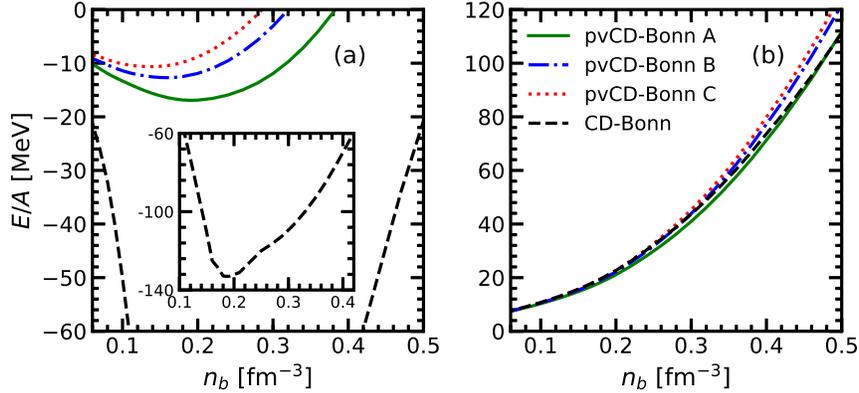}\\
	\caption{The equations of state of symmetric matter and pure neutron matter from CD-Bonn and pvCD-Bonn potentials.}
	\label{CDbonnEOS}
\end{figure}

The pseudoscalar (PS) coupling and pseudovector (PV) coupling schemes between pion and nucleon from quantum field theory are equivalent for on-shell nucleon, since their coupling constants satisfy the relation, $g_\pi/2M=f_\pi/m_\pi$. However, their off-shell matrices have significant differences as shown in our previous work about pvCD-Bonn potentials~\cite{wang19}. In present relativistic nuclear many-body methods, the PV coupling is adopted, which can  suppress the contributions from the antinucleon, i.e., pair suppression mechanism, and generate reasonable physical results. On the contrary, the PS coupling will drive a largely spurious attraction~\cite{serot86,fuchs98}. Therefore, in relativistic Hartree-Fock model~\cite{bouyssy87,long06,long07} and RBHF model, the PV coupling interaction between pion and nucleon is required. In Fig.~\ref{CDbonnEOS}, the equations of state of symmetric nuclear matter and pure neutron matter from RBHF model within the original CD-Bonn potential by Machleidt~\cite{machleidt01} are plotted and are compared with the results within pvCD-Bonn potentials. For symmetric nuclear matter, the saturation binding energy of CD-Bonn potential is about $-140$ MeV. This extra attraction is obviously derived from the PS coupling. On the other hand, the equation of state of pure neutron matter from CD-Bonn potential is quite similar with those from pvCD-Bonn potentials. It is because that the contribution from pion interaction in one-boson-exchange potential is largely suppressed in pure neutron matter due the total isospin of two neutrons, $T=1$, which almost does not play any role in total energy.  

\begin{figure}[http]
                    \centering
                    \includegraphics[scale=0.6]{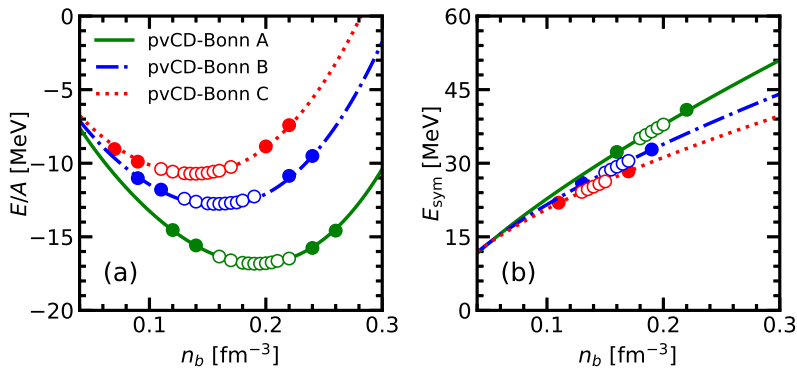}\\
                    \caption{The energies per nucleon and symmetry energies parameterized by Eqs. (\ref{FunEOS}) and (\ref{Esym}) for 
                    pvCD-Bonn potentials. The opened circles are the fitting data and solid ones are used to check the reliability of parameterizations.}
                    \label{FunEOSf}
                \end{figure}

The calculations of RBHF model are usually complicated and time-consuming. To apply these results in other aspects easily, the equations of state of symmetric nuclear matter from pvCD-Bonn A, B, C potentials are better to be parameterized around the saturation density, $n_0$. It is suggested that the energy per nucleon can be expanded as the following function in Ref.~\cite{gandolfi12},  
                \begin{equation}\label{FunEOS}
                    \frac{E_0}{A}(n_b)=a\left(\frac{n_b}{n_0}\right)^\eta + b\left(\frac{n_b}{n_0}\right)^\beta~~~~~(\alpha_J=0).
                \end{equation}
Furthermore, the symmetry energy closed to the saturation density $n_0$ also can be expressed as 
                \begin{equation}
                    E_\mathrm{sym}(n_b)=c\left(\frac{n_b}{n_0}\right)^\gamma.
                \end{equation}
It can be approximately obtained from differences between the energies per nucleon of pure neutron matter and symmetric nuclear matter,
             \begin{equation}\label{Esym}
              E_\mathrm{sym}(n_b) \approx \frac{E}{A}(n_b,\alpha=1)-\frac{E}{A}(n_b,\alpha=0).
             \end{equation} 
The corresponding values of $a~,b,~c$ and $\eta,~\beta,~\gamma$ are obtained by fitting the numerical results of RBHF model with pvCD-Bonn A, B, C potentials, which are listed in Table~\ref{ParamEOS} and shown in Fig.~\ref{FunEOSf}. These parameters are also consistent with the results from Bonn potentials worked out by Tong {\it et al.}~\cite{tong18}  It can be found that the energy per nucleon and symmetry energy from RBHF model shown as open and solid circles are well parameterized by the fitting functions, Eqs. (\ref{FunEOS}) and (\ref{Esym}), denoted by the solid curves in Fig.~\ref{FunEOSf}.              
                                             
                \begin{table}[http]
                    \centering 
                    \caption{Parameters in Eqs. (\ref{FunEOS}) and (\ref{Esym}) for the equations of state of symmetric nuclear matter and symmetry energies from RBHF model with pvCD-Bonn A, B, C potentials.}
                    \label{ParamEOS}
                    \setlength{\tabcolsep}{1.2mm}{
                    \begin{tabular}{cccccccc}
                    \toprule 
                                 & $n_0$ [fm$^{-3}$] & $a$~[MeV]& $\eta$ & $b$~[MeV] & $\beta$ & $c$~[MeV] & 
                                 $\gamma$ \\
                                 \hline 
                                 &       &          &         &         &         &        &      \\
                     pvCD-Bonn A & 0.192 & -20.86   &   0.64  &   4.03  & 3.28    & 36.75  & 0.73 \\
                                 &       &          &         &         &         &        &      \\
                     pvCD-Bonn B & 0.158 & -15.87   &   0.58  &   2.96  & 3.08    & 29.05  & 0.65 \\
                                 &       &          &         &         &         &        &      \\
                     pvCD-Bonn C & 0.139 & -13.06   &   0.52  &   2.34  & 2.95    & 25.12  & 0.59 \\
                    \hline
                    \hline
                    \end{tabular}}
                \end{table}
                       
The saturation properties, saturation density, $n_0$ and corresponding energy per nucleon, $E/A$, incompressibility, $K_\infty$, symmetry energy, $E_\mathrm{sym}$, the slope of symmetry energy, $L$ are summarized in Table~\ref{SatProperties} for pvCD-Bonn A, B, C potentials from RBHF model. The results from Bonn potentials are also listed for comparison worked by Tong {\it et al}.~\cite{tong18} These empirical observables are strongly correlated to tensor components in $NN$ potentials, i.e., the $D$-state probability of deuteron, $P_D$. The incompressibilities, symmetry energies, and their slopes at saturation densities satisfy the conventional constraints extracted from properties of finite nuclei within limits. Especially, the smaller values, $L$, are preferred by recent measurements about the neutron skin of finite nuclei and gravitational waves from binary neutron star merger~\cite{abbott17a,abbott18a}. Although $G$ matrices were obtained by Thompson equation for Bonn potentials, their $P_D$ dependence of saturation properties are accordance with those derived by BbS equation for pvCD-Bonn potentials. 

                \begin{table}[http]
                \centering
                \caption{The saturation properties of nuclear matter. The data of Bonn A, B, C are collected 
                from~\cite{tong18}. $n_0$ refers to the saturation densities.}
                \label{SatProperties}
                \setlength{\tabcolsep}{1.0mm}{
                    \begin{tabular}{lcccccccc}
                    \toprule
                               & $P_D$  &$n_0$& $E/A$&
                               $K_\infty$ &  $E_\mathrm{sym}$ & $L$ \\
                    \hline
                               &        &         &       &       &        &       &       &      \\
                    pvCD-Bonn A&4.22$\%$& 0.192   &-16.83 & 315   & 36.8   &  80.5  \\
                               &        &         &       &       &        &       &       &      \\
                    pvCD-Bonn B&5.45$\%$& 0.158   &-12.91 & 206   & 29.1   &  56.7  \\
                               &        &         &       &       &        &       &       &      \\
                    pvCD-Bonn C&6.05$\%$& 0.139   &-10.72 & 151   & 25.1   &  44.5  \\
                               &        &         &       &       &        &       &       &      \\
                    Bonn A     &4.47$\%$& 0.180   &-15.38 & 286   & 33.7   &  75.8  \\
                               &        &         &       &       &        &       &       &      \\
                    Bonn B     &5.10$\%$& 0.164   &-13.44 &  222  & 29.9   &  63.0  \\
                               &        &         &       &       &        &       &       &      \\
                    Bonn C     &5.53$\%$& 0.149   &-12.12 & 176   & 26.8   &  51.7  \\
                               &        &         &       &       &        &       &  \\

                    Empirical  &  &0.16$\pm$0.01  &-16$\pm$1
                               &240$\pm$20&31.7$\pm$3.2
                    &  58.7$\pm$ 28.1  \\
                           &  &  &\cite{danielewicz09}
                    &\cite{garg18}&\cite{oertel17} 
                    & ~\cite{oertel17}   \\
                    \hline
                    \hline
                    \end{tabular}}
                \end{table}

\begin{figure}[http]
                    \centering
                    \includegraphics[scale=0.6]{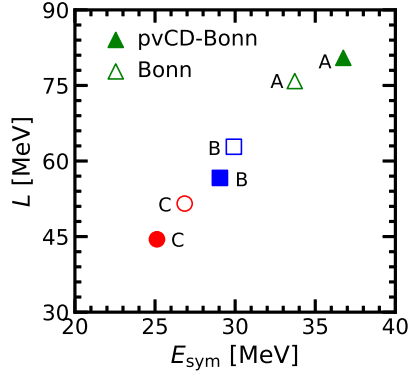}\\
                    \caption{The relations between symmetry energies  and their slopes at saturation densities for pvCD-Bonn and Bonn potentials.}
                    \label{Esym0vsL0}
                \end{figure}

In nuclear density functional theories, it was found that the slope of symmetry energy at saturation density has strong linear correlations with the neutron skins of $^{208}$Pb and the symmetry energy~\cite{li08,li19}. In Fig.~\ref{Esym0vsL0}, the relation between $E_\mathrm{sym}$ and $L$ at saturation density are shown for pvCD-Bonn potentials and Bonn potentials. They also have the strong linear correlation with different $P_D$. In general, the lower $P_D$ provides a larger symmetry energy and a larger slope. Since the tensor force will suppress the depth of bound state in symmetric nuclear matter. Recently, the behaviors of symmetry energy at high density also attracted the wide attentions. The ASY-EOS experiment at GSI in 2016 showed that the symmetry energy at $2n_{E0}$ and  $3n_{E0}$ should be around $50.82-60.39$ MeV and $64.34-84.74$ MeV, respectively, where $n_{E0}=0.16$ fm$^{-3}$ is the empirical saturation density~\cite{russotto16}. Furthermore, many theoretical works also presented their constraints on the density dependence of symmetry energy, like the chiral effective theory~\cite{holt17,lim19}. Therefore, in Table~\ref{Esym(nb)},  symmetry energies at $n_{E0}$, $2n_{E0}$, and $3n_{E0}$ are listed for pvCD-Bonn and Bonn potentials. It is obvious that the density-dependent behaviors of symmetry energy from pvCD-Bonn A and Bonn A potentials satisfy the observations from the ASY-EOS experiment.
\               
                 
                \begin{table}[http]
                    \centering 
                    \caption{The values of symmetry energy at different densities obtained from pvCD-Bonn and Bonn potentials. $n_b$ in unit fm$^{-3}$ and symmetry energy in unit MeV.}
                    \label{Esym(nb)}
                    \setlength{\tabcolsep}{1.2mm}{
                    \begin{tabular}{c|cccccc}
                    \toprule 
                        $n_b$   & pvCD-Bonn A& pvCD-Bonn B& pvCD-Bonn C&  Bonn A& Bonn B & Bonn C \\  
                          \hline 
                                 &        &          &         &         &         &        \\
                     $0.16$      & 32.17 &  29.29   &  27.29  &  30.87  & 29.41   &  28.10  \\
                                 &       &          &         &         &         &           \\
                     $0.32$      & 53.36 &  45.96   &   41.08  & 51.92  &  47.77  & 43.79 \\
                                 &       &          &         &         &         &            \\
                     $0.48$     & 71.74 & 59.82  &  52.19  &  70.37 & 63.45    & 56.77  \\
                    \hline
                    \hline
                    \end{tabular}}
                \end{table}

\subsection{The  potentials of symmetric nuclear matter}
\begin{figure}[http] 
            	\centering
            	\includegraphics[scale=0.6]{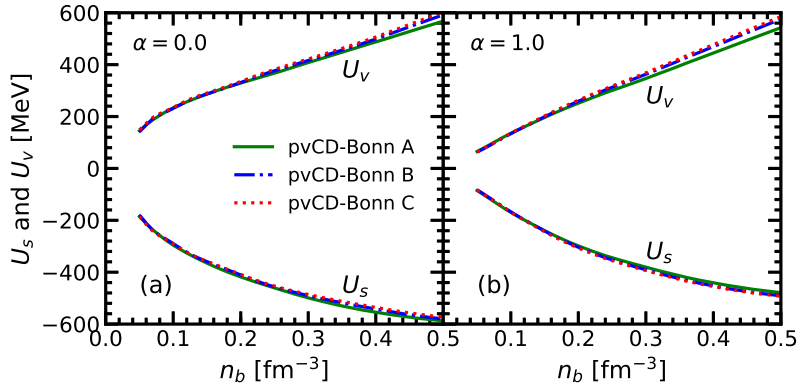}\\
            	\caption{The scalar and vector potentials as functions of nucleon density in symmetric nuclear matter and pure neutron matter.}
            	\label{UsUv}
            \end{figure}

The scalar and vector potentials are two important quantities in RBHF model to connect the Dirac equation and $G$ matrices through the nucleon single-particle potential, which also denote the attraction and repulsion of $NN$ interaction at different ranges, respectively. In Fig.~\ref{UsUv}, the scalar and vector potentials from pvCD-Bonn potentials in symmetric nuclear matter and pure neutron matter are given in panel (a) and panel (b), respectively.  In present work, they are assumed to be only density dependent and momentum independent, which are extracted from Eqs.~\eqref{SP-Pot} and \eqref{HF-SP-Pot}. The self-consistent calculation of RBHF model is determined by the convergence of proton and neutron scalar potentials. At low density regions, $U_S$ and $U_V$ from three pvCD-Bonn potentials are almost identical. Their differences among the three potentials become obvious with density increasing. For vector potential, the pvCD-Bonn C provides more repulsive contribution both in symmetric nuclear matter and pure neutron matter, while for scalar potential, pvCD-Bonn C potential generates more attractive component in pure neutron matter case. 
        
\begin{figure}[http]
	\centering
	\includegraphics[scale=0.6]{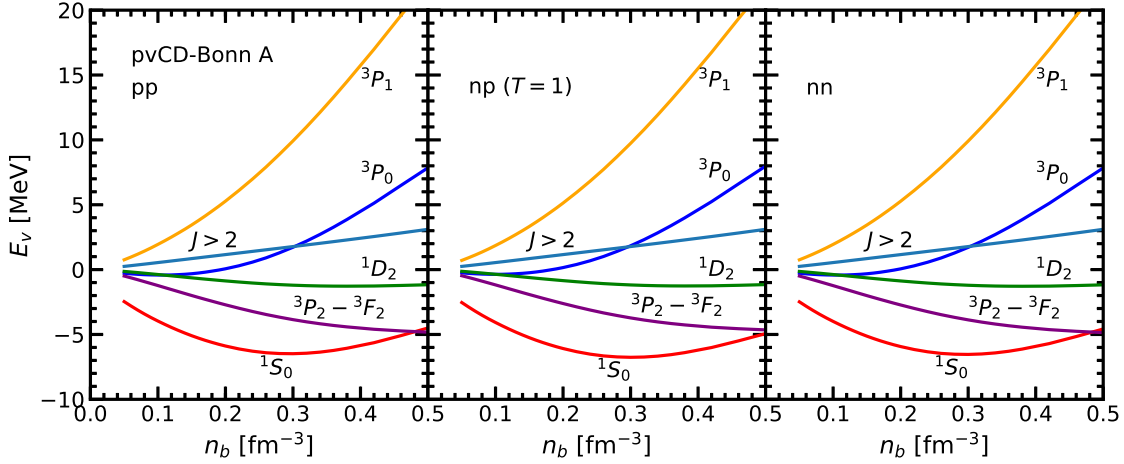}\\
	\caption{Partial wave contributions from isospin-triplet channels to the potential energy in symmetric matter obtained by pvCD-Bonn A.}\label{PWpvA1}
\end{figure}

 The BbS equation actually was solved in partial wave $LSJ$ representation. In these calculations, the largest total angular momentum is taken up to $J=8$.  In Fig.~\ref{PWpvA1}, the main contributions of partial waves to the potential energy of nucleon at isospin-triplet channels [$pp,~nn,~np(T=1)$] are displayed for the symmetric nuclear matter from pvCD-Bonn A potential. There are quite small differences among $pp,~nn,~np(T=1)$, related to the charge symmetry breaking (CSB) and charge independent breaking (CIB) effects of realistic $NN$ potential~\cite{machleidt01}. These two effects will be discussed in detail later. The partial wave contributions with $J\leqslant2$ play the dominant roles in the potential energy of nucleon. At low density region, $^1S_0$ channel generates most of the attraction, which represents the central force in $NN$ interaction. Furthermore, $^1D_2$ and $^3P_2$-$^3F_2$ channels also provide the partially bound energies. On the other hand, $^3P_0$ and $^3P_1$ channels give the repulsive contributions, where $^3P_1$ channel has the stronger magnitude.
 
\begin{figure}
	\centering
	\includegraphics[scale=0.6]{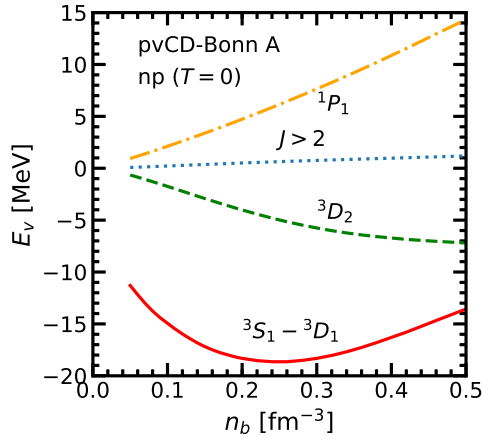}\\
	\caption{Partial wave contributions from isospin-singlet channel to the potential energy in symmetric matter obtained by pvCD-Bonn A potential.}\label{PWpva0}
\end{figure}

In Fig.~\ref{PWpva0}, the corresponding partial wave contributions from isospin-singlet channel to the potential energy is also shown. There is only $np$ potential due to the Pauli exclusion principle. The spin-triplet channel, $^3S_1$-$^3D_1$,  provides the strongest attractive contribution and $^3P_1$ channels generates the largest repulsive one. Especially, the $^3S_1$-$^3D_1$ coupled channel mainly comes from the tensor force in $NN$ potentials. There is a saturation point in the contribution of $^3S_1$-$^3D_1$ channel, whose saturation density is very closed to that of symmetric matter. In fact, it can be found the contributions of each partial wave from three pvCD-Bonn potentials are almost the same except the one from $^3S_1$-$^3D_1$ channel. Therefore, the energy from  $^3S_1$-$^3D_1$ channel determines the saturation properties of symmetric nuclear matter. This is why the saturation properties of nuclear matter is so closely related with the strength of tensor force, or alternatively, the $D$-state probability of deuteron, $P_D$~\cite{hu13}. 

To discuss the CSB and CIB effects of $NN$ potentials in nuclear matter, the detailed values of partial wave contributions in symmetric nuclear matter from pvCD-Bonn potentials are listed in Tables~\ref{tab-pwa1} and \ref{tab-pwa2} at empirical saturation density, $n_{E0}=0.16$ fm$^{-3}$ and $n_b=0.32$ fm$^{-3}$, respectively. These partial wave contributions from pvCD-Bonn A, B, C potentials are very similar in each channel, expect the coupled channels,  $^3S_1$-$^3D_1$ and $^3P_2$-$^3F_2$. It is easy to understand these results, because pvCD-Bonn potentials can precisely describe the phase shifts obtained from the Nijmegen partial wave analysis. Their differences mainly comes from the mixing parameters, $\varepsilon_1$ and $\varepsilon_2$ of $^3S_1$-$^3D_1$ and $^3P_2$-$^3F_2$ channels, which also are determined by the tensor force component of $NN$ potential 

In symmetric nuclear matter, the Fermi momenta of proton and neutron are completely identical. The differences of partial wave contribution in the same channel can be derived only from the CSB or CIB effect. The charge symmetry of $NN$ potential is invariant under a transformation from proton-proton ($pp$) interaction to neutron-neutron ($nn$) interaction after removing the Coulomb force between protons. The CSB effect in nuclear matter is regarded to the differences between energy contributions from $pp$ and $nn$ interactions. From Tables ~\ref{tab-pwa1} and \ref{tab-pwa2}, it demonstrates that the CSB effect is mainly embodied in $^1S_0$ channel. The energy differences from $nn$ and $pp$ potentials are about $0.04-0.06$ MeV. In addition, the CIB effect is obtained by comparing the partial wave contributions from $np$ with those from $pp$ and $nn$. In each isospin-triplet channel, the potential energy from $np$ interaction has the significant distinction with those from $pp$ and $nn$ interactions. The largest difference is around $0.35$ MeV. Therefore, in nuclear matter with pvCD-Bonn potentials, the CIB effect has a more obvious signature comparing to the CSB effect. Actually, there was also a similar conclusion for the $NN$ singlet scattering length and effective range at $^1S_0$ channel in Refs.~\cite{wiringa95,machleidt01}.

\begin{table}[thbp]
	\centering
	\caption{The partial wave contributions derived by $nn, pp, np$ interactions from pvCD-Bonn potentials at $n_b=0.16$ fm$^{-3}$ in the unit MeV.}
	\label{tab-pwa1}
	\setlength{\tabcolsep}{1.2mm}{
		\begin{tabular}{r|rrrr|rrrr|rrrr}
			\toprule
			&      &      &      &       &      &      &      &       &      &      &      &       \\
			& \multicolumn{4}{c|}{pvCD-Bonn A}
			& \multicolumn{4}{c|}{pvCD-Bonn B}
			& \multicolumn{4}{c} {pvCD-Bonn C} \\
			&      &      &      &       &      &      &      &       &      &      &      &       \\
			\hline
			& $pp$ & $np$ & $nn$ & total & $pp$ & $np$ & $nn$ & total & $pp$ & $np$ & $nn$ & total \\
			$^1S_0$&-5.28 &-5.44 &-5.32 &-16.04 &-5.28 &-5.46 &-5.32 &-16.06 &-5.26 &-5.45 &-5.30 &-16.01 \\
			$^3P_0$&-0.26 &-0.18 &-0.27 & -0.71 &-0.31 &-0.23 &-0.31 &-0.85  &-0.32 &-0.24 &-0.32 &-0.88  \\
			$^1P_1$&      & 3.64 &      &  3.64 &      & 3.77 &      & 3.77  &      & 3.78 &      & 3.78  \\
			$^3P_1$& 3.78 & 3.66 & 3.78 & 11.22 & 3.85 & 3.69 & 3.84 & 11.38 & 3.89 & 3.78 & 3.89 & 11.56 \\
			$^3S_1-$$^3D_1$
			&      &-17.50&      &-17.50 &      &-14.72&      &-14.72 &      &-12.72&      &-12.72 \\
			$^1D_2$&-0.68 &-0.66 &-0.69 & -2.03 &-0.67 &-0.65 &-0.68 &-2.00  &-0.67 &-0.65 &-0.68 &-2.00  \\
			$^3D_2$&      &-3.15 &      & -3.15 &      &-3.16 &      &-3.16  &      &-3.17 &      &-3.17  \\
			$^3P_2-$$^3F_2$
			&-2.14 &-2.07 &-2.15 & -6.46 &-2.06 &-1.98 &-2.08 &-6.22  &-2.05 &-1.98 &-2.06 &-6.19  \\
			$^1F_3$&      & 0.65 &      &  0.65 &      & 0.65 &      & 0.65  &      & 0.64 &      & 0.64  \\
			$^3F_3$& 0.43 & 0.37 & 0.43 &  1.23 & 0.43 & 0.37 & 0.43 & 1.23  & 0.42 & 0.37 & 0.42 & 1.21  \\
			$^3D_3-$$^3G_3$
			&      & 0.18 &      &  0.18 &      & 0.16 &      & 0.16  &      & 0.15 &      & 0.15  \\
			\hline
			\hline
	\end{tabular}}
\end{table}

\begin{table}[thbp]
	\centering
	\caption{The partial wave contributions derived by $nn, pp, np$ interactions from pvCD-Bonn potentials at $n_b=0.32$ fm$^{-3}$ in the unit MeV.}
	\label{tab-pwa2}
	\setlength{\tabcolsep}{1.2mm}{
		\begin{tabular}{r|rrrr|rrrr|rrrr}
			\toprule
			&      &      &      &       &      &      &      &       &      &      &      &       \\
			& \multicolumn{4}{c|}{pvCD-Bonn A}
			& \multicolumn{4}{c|}{pvCD-Bonn B}
			& \multicolumn{4}{c}{pvCD-Bonn C} \\
			&      &      &      &       &      &      &      &       &      &      &      &       \\
			\hline
			& $pp$ & $np$ & $nn$ & total & $pp$ & $np$ & $nn$ & total & $pp$ & $np$ & $nn$ & total \\
			$^1S_0$&-6.45 &-6.74 &-6.50 &-19.69 &-6.54 &-6.88 &-6.60 &-20.02 &-6.54 &-6.89 &-6.60 &-20.03 \\
			$^3P_0$& 2.20 & 2.31 & 2.19 &  6.70 &1.95  & 2.05 & 1.94 & 5.94  & 1.83 & 1.92 & 1.81 & 5.56 \\
			$^1P_1$&      & 8.89 &      &  8.89 &      &  9.10&      & 9.10  &      & 9.05 &      & 9.05  \\
			$^3P_1$&11.00 &10.80 & 10.98& 32.78 &11.32 & 10.96& 11.29&33.57  &11.47 & 11.29& 11.49& 34.25 \\
			$^3S_1-$$^3D_1$
			&      &-19.44&      &-19.44 &      &-11.67&      &-11.67 &      &-7.18 &      &-7.18 \\
			$^1D_2$&-1.24 &-1.22 &-1.24 & -3.70 &-1.21 &-1.19 &-1.22 &-3.62  &-1.23 &-1.21 &-1.24 &-3.68  \\
			$^3D_2$&      &-6.01 &      & -6.01 &      &-6.22 &      &-6.22  &      &-6.25 &      &-6.25  \\
			$^3P_2-$$^3F_2$
			&-4.03 &-3.89 &-4.05 &-11.97 &-3.92 &-3.77 &-3.94 &-11.63 &-3.93 &-3.79 &-3.94 &-11.66 \\
			$^1F_3$&      & 1.46 &      &  1.46 &      & 1.46 &      & 1.46  &      & 1.45 &      & 1.45  \\
			$^3F_3$& 1.07 & 0.97 & 1.07 &  3.11 &1.07  & 0.98 & 1.07 & 3.12  & 1.08 & 0.98 & 1.08 & 3.10  \\
			$^3D_3-$$^3G_3$
			&      & 0.54 &      &  0.54 &      & 0.50 &      & 0.50  &      & 0.48 &      & 0.48  \\
			\hline
			\hline
	\end{tabular}}
\end{table}        
            
\subsection{The properties of asymmetric nuclear matter}
\begin{figure}[http]
                \centering
                \includegraphics[scale=0.6]{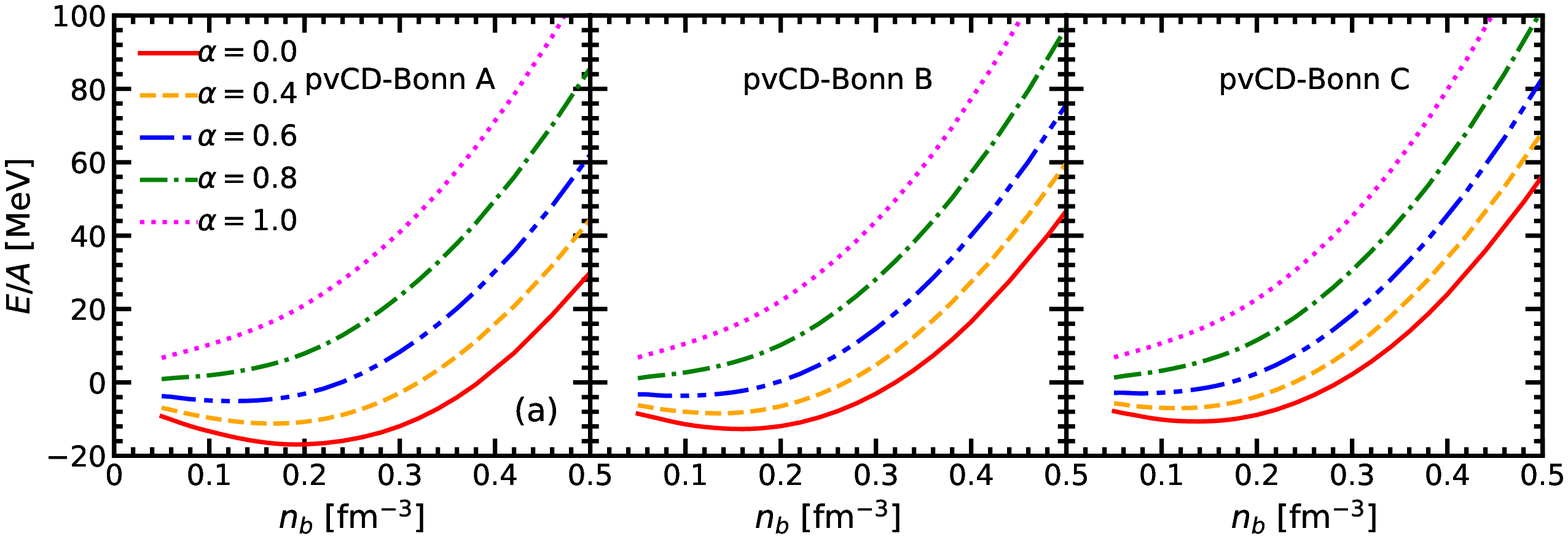}\\
                 \caption{\small{The equations of state of asymmetric nuclear matter calculated by RBHF model with different $\alpha$ from pvCD-Bonn A, B, C potentials.}}
                \label{Asymmatter}
            \end{figure}

The asymmetric nuclear matter with different fractions of protons and neutrons are very important for the investigations of compact star and supernova simulations~\cite{shen98a,shen98b,tong19}. The Pauli operators in the medium BbS equation will become more complicated due to the distinguished Fermi integration spheres of proton and neutron. The detailed formulas about the evaluation of asymmetric nuclear matter are given in  the appendix B. The equations of state of asymmetric nuclear matter with different asymmetry parameters, $\alpha$, are plotted in Fig.~\ref{Asymmatter} from pvCD-Bonn potentials. With the neutron numbers increasing, the equation of states of asymmetric nuclear matter are not saturated and not bound above $\alpha=0.8$. The differences of three equations of state among pvCD-Bonn A, B, C potentials also quickly reduced for larger $\alpha$. 
 
\begin{figure}[h]
                \centering
                \includegraphics[scale=0.6]{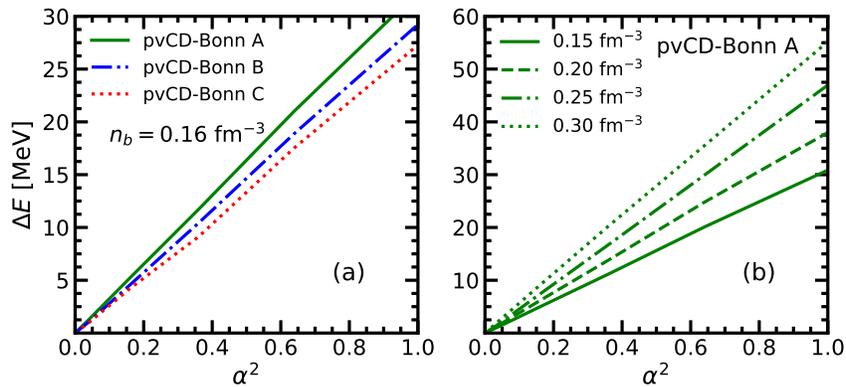}\\
                \caption{The energy differences $\Delta E$ at fixed baryon density as functions of $\alpha^2$. In panel (a), the $\Delta E$ from pvCD-Bonn A, B, C potentials at fixed density $n_{E0} = 0.16$ fm$^{-3}$. In panel (b), $\Delta E$ from pvCD-Bonn A potential at $n_b = 0.15,~0.20,~0.25,~0.30$ fm$^{-3}$.}
                \label{DE}
            \end{figure}

The energy per nucleon in asymmetric nuclear matter is regarded to be expanded as a polynomial with a variable $\alpha^2$ around $\alpha=0$,       
            \begin{equation} \label{E/A-Expansion}
                \frac{E}{A}(n_b,\alpha)=  \frac{E_0}{A}(n_b) +  E_\text{sym}(n_b)\alpha^2 + 
                \mathcal{O}(\alpha^4),
            \end{equation}
where the coefficient in second term, $E_\text{sym}(n_b)$ is defined as the symmetry energy. In Fig.~\ref{DE}, the energy differences $\Delta E = \frac{E}{A}(n_b,\alpha) - \frac{E}{A}(n_b,0)$, as functions of $\alpha^2$ are plotted in present calculations. It is a suitable way to check the expansion convergence of $\alpha$ in Eq.~\ref{E/A-Expansion}. In the panel (a) of this figure, energy differences from three pvCD-Bonn potentials at empirical saturation density $n_{E0}=0.16$ fm$^{-3}$ have almost the linear relations with $\alpha^2$. It demonstrates that the neglect of higher terms about $\alpha^2$ in the expansion of asymmetric nuclear matter is reasonable. In panel~(b), the validity of such linear relation is checked at different baryon densities $n_b = 0.15,~0.20,~0.25,~0.30$ fm$^{-3}$ with pvCD-Bonn A potential, which still work well.
           
            \begin{figure}
              \centering
              \includegraphics[scale=0.6]{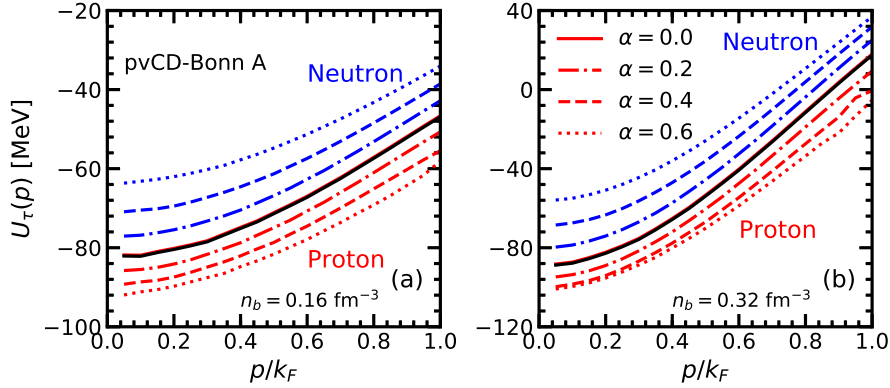}\\
              \caption{The neutron (upper) and proton (lower) single-particle potentials in asymmetric nuclear matter at different asymmetry parameter obtained by pvCD-Bonn A potential.}\label{UsppvA}
            \end{figure}
            
In Fig.~\eqref{UsppvA}, the neutron and proton single-particle potentials as functions of momentum are displayed with different asymmetric parameters at $n_{E0}=0.16$ fm$^{-3}$ (panel (a)) and $n_b=0.32$ fm$^{-3}$ (panel (b)). In symmetric nuclear matter, these potentials for neutron and proton are identical. With the fractions of neutron increasing, the neutron single-particle potential become more repulsive, while the case of proton is opposite. It means that the proton obtains more attractive contribution from the $NN$ potential. Therefore the effective neutron mass is larger than the proton one in neutron-rich matter.  The differences between neutron and proton single-particle potentials become smaller with momentum for a fixed $\alpha$ and increase with the nucleon density. This behavior of nucleon single-particle potential is completely consistent with the work using the Bonn potentials by Sammarruca~\cite{sammarruca10}.

   \section{summary and outlook} \label{End}
Properties of nuclear matter were investigated in relativistic Brueckner-Hartree-Fock (RBHF) model with the latest charge-dependent nucleon-nucleon potentials, pvCD-Bonn A, B, C, where the coupling scheme between pion and nucleon is taken as pseudovector form. These three potentials have different tensor components.  Furthermore, the center-of-mass momentum related to $G$ matrix was exactly integrated in present work without the conventional angle-averaged approximation.  

Firstly, the equations of state of symmetric nuclear matter and pure neutron matter with three pvCD-Bonn potentials were obtained. Their saturation densities and saturation energies were closed to the empirical data for symmetric nuclear matter. These saturation properties are strongly related to the tensor components of $NN$ potentials, which can be presented by the $D$-state probability of deuteron, $P_D$. Generally speaking, the smaller tensor component provides larger saturation density and more attractive binding energy. They could be summarized as a relativistic Coester band with the results from pvCD-Bonn and Bonn potentials. Furthermore, they were also compared to the results from nonrelativistic framework. The relativistic effect provides more repulsive contribution and generates reasonable saturation properties. For the pure neutron matter, the equations of state from pvCD-Bonn potentials almost were identical, since the tensor contribution is very weak in the isopin $T=1$ case.

The original CD-Bonn potential with pseudoscalar (PS) coupling was also applied to calculate the properties of nuclear matter. It was confirmed that the PS coupling between pion and nucleon provides a too much attractive contribution and generates over-bound state for symmetric nuclear matter, while it did not influence the pure neutron matter. With these equations of state, the additional properties, such as incompressibility, symmetry energy and its slope at saturation density were evaluated. These properties also satisfied the constraints extracted from the finite nuclei experiments. The symmetry energies at higher nuclear densities, such as twice or three times empirical saturation density from pvCD-Bonn potentials were accordance with recent results from ASY-EOS experiment at GSI laboratory. 

Through discussing the partial wave contributions to potential energy, it was found that the differences among three pvCD-Bonn potentials for the symmetric nuclear matter mainly came from the coupled channel $^3S_1$-$^3D_1$, since their phase shifts were only distinguished by the mixing parameters $\varepsilon_1$. In addition, the charge symmetry breaking (CSB) and charge independent breaking (CIB) effects in nuclear matter were also investigated. The CSB effect derived by the difference between $pp$ and $nn$ potentials embodied in $^1S_0$  channel about $0.04-0.06$ MeV. The CIB effect from $np$ to $pp$ or $nn$ potentials appeared in each isospin-triplet channel and was more obvious than the CSB effect. The equations of state of asymmetric nuclear matter were also calculated with pvCD-Bonn potentials. They were not bound together when the asymmetry parameters were larger than $0.8$. The magnitude of neutron single-particle potential was higher than that from proton in neutron-rich matter, which leads to the fact that the neutron effective mass in nuclear medium is larger than the proton one. 
    
The RBHF model is a very powerful {\it ab initio} method in relativistic framework, which can explain the saturation properties of nuclear matter reasonably by using only two-body $NN$ potential. With the newly developed high-precision charge-dependent Bonn potentials, pvCD-Bonn A, B, C, more investigations will be done in nuclear physics, such as properties of neutron star, the superfluity of nucleon in medium, and the saturation mechanism of nuclear matter in future.

\section*{Acknowledgments}
This work was supported in part by the National Natural Science Foundation of China (Grant No. 11775119, No. 11405116, and No. 11675083), the Natural Science Foundation of Tianjin, and China Scholarship Council (Grant No. 201906205013 and No. 201906255002).

   \appendix\label{Appendix}
        \section{In-medium Blanckenbecler-Sugar equation}
In the conventional RBHF model, the $G$ matrix was solved via the in-medium Thompson equation, since the Bonn potentials were obtained by fitting the $NN$ scattering data with Thompson equation in free space~\cite{machleidt89,brockmann90}. The pvCD-Bonn potentials were generated by Blanckenbecler-Sugar (BbS) equation to keep the consistency with the original CD-Bonn potential. Therefore, the in-medium BbS equation must be derived in this work, which has been mentioned in the appendix A of Ref.~\cite{brockmann90}. The procedure of nucleon-nucleon scattering  is dominated by Bethe-Salpter (BS) equation, which is written as
            \begin{equation}\label{BS-Eq}
            \mathcal{T} = \mathcal{V}+\mathcal{V}\mathcal{G}\mathcal{T},
            \end{equation}
where $\mathcal{T}$ is an invariant amplitude for $NN$ scattering and $\mathcal{G}$ is a two-nucleon propagator. The BS equation in Eq.~\eqref{BS-Eq} is defined in four-dimension space, explicitly, at center-of-mass frame,
            \begin{equation}\label{BS-4D}
            \mathcal{T}(q',q|P)= \mathcal{V}(q',q|P) +
            \int \frac{d^4k}{(2\pi)^4}\mathcal{V}(q',k|P)
            \mathcal{G}(k|P)\mathcal{T}(k,q|P),
            \end{equation}
where $q,~k,~q'$ are initial, intermediate, and final relative four-momenta, respectively. $P=(p_1+p_2)/2$ is one half of total momentum.

The propagator, $\mathcal{G}$ in Eq.\eqref{BS-Eq} is given as
            \begin{equation}\label{BS-Prop}
            \mathcal{G}(k|P) = \frac{i}{  p_1\!\!\!\!\!\slash -M_1+i\epsilon}
            \frac{i}{p_2\!\!\!\!\!\slash-M_2+i\epsilon},
            \end{equation}
where $M_1$ and $M_2$ denote the nucleon masses, $p_1=(E_1, \mathbf{p}_1)$ and $p_2=(E_2, \mathbf{p}_2)$ are four-momenta of two one-shell nucleons, respectively.

Actually, the BS equation is very difficult to be solved in four-dimension space for numerical calculation. To simplify BS equation, the propagator $\mathcal{G}$ must be reduced into three-dimension space as $g$. It should reserve unitary and covariant properties of original $\mathcal{G}$ in this process. Therefore, $g$ and $\mathcal{G}$ should have the same discontinuity across the branch cut,
            \begin{equation}
            \text{Im}g(k|P)=\text{Im}\mathcal{G}(k|P)=-2\pi^2
            (p_1\!\!\!\!\!\slash+M_1) (p_2\!\!\!\!\!\slash+M_2)
            \delta^{(+)}(p_1^2-M_1^2)\delta^{(+)}(p_2^2-M_2^2),
            \end{equation}
where $\delta^{(+)}(p_i^2-M_i^2)$ means that only the on-shell nucleons are involved (exclusion of anti-nucleon).

It is more convenient to express $\text{Im} g$ in center-of-mass frame,
            \begin{align}
            \nonumber
            \text{Im}g(k|P)&=-2\pi^2(p_1\!\!\!\!\!\slash+M_1)
            (p_2\!\!\!\!\!\slash+M_2)\frac{\delta(p_1^0-E_1)
            \delta(p_2^0-E_2)}{4E_1E_2}\\
            \nonumber
            &=-2\pi^2\frac{M_1M_2}{E_1E_2} \Lambda_+^{(1)}(\mathbf{p}_1)
            \Lambda_+^{(2)}(\mathbf{p}_2)\delta(2P_0-W_k) \delta(k_0 -E_1/2 +E_2/2) \\
            \label{BS-Prop-3D-Im}
            &=-4\pi^2W_k\frac{M_1M_2}{E_1E_2} \Lambda_+^{(1)}(\mathbf{p}_1)
            \Lambda_+^{(2)}(\mathbf{p}_2)\delta(s'-W_k^2+4\mathbf{P}^2)
            \delta(k_0 -E_1/2 +E_2/2)
            \end{align}
with $W_k = E_1+E_2$ and the immediate total energy, $s' = 4P_0^2-4\mathbf{P}^2$. The projection operator is defined by
            \begin{equation}\label{Proj-Spinor}
            \Lambda_+^{(i)}(p) = \frac{p\!\!\!\slash+M_i}{2M_i} =
            \sum_{\lambda}u_i(\mathbf{p},\lambda)
            \bar{u}_i(\mathbf{p},\lambda),
            \end{equation}
where $\lambda$  represents  the eigenvalue of spin operator. Furthermore, the imaginary part \eqref{BS-Prop-3D-Im}, $g(k,s)$ can be constructed by the dispersion integral,
            \begin{align}\label{Dis-Int-BbS}
            g(k,s) =\frac{1}{\pi} \int_{4M^2}^{+\infty}
            \frac{ds'}{s'-s-i\epsilon}\text{Im}g(k,s').
            \end{align}
Here, we consider that the starting energy is written as $W_0 = E_{\mathbf{P}+\mathbf{q}} + E_{\mathbf{P}-\mathbf{q}}$ and $s=W_0^2 -4\mathbf{P}^2$. Therefore, the four-dimension propagator, \eqref{BS-Prop} can be reduced to three-dimension one as
            \begin{equation}\label{BS-Prop-3D}
            G(k|P)\rightarrow g(k,s) =2\pi \delta(k_0-E_1/2+E_2/2)\bar{g}(\mathbf{k},s)
            \end{equation}
with different choices for $\bar{g}(\mathbf{k},s)$. After integrating \eqref{Dis-Int-BbS}, the BbS propagator is obtained explicitly~\cite{blankenbecler66},
            \begin{equation}\label{BbS-Prop}
            \bar{g}_\text{BbS}(\mathbf{k},s)= 2W_k\frac{M_1M_2}{E_1E_2}
            \frac{\Lambda^{(1)}_+(\mathbf{P}+\mathbf{k})
            \Lambda^{(2)}_+(\mathbf{P}-\mathbf{k})}
            {W_0^2-W_k^2+i\epsilon}.
            \end{equation}
 The in-medium effects must be taken into account in the BbS propagator for nuclear many-body system with Pauli operator $Q$,
            \begin{equation}\label{BbS-Prop-IM}
            \bar{g}_\text{BbS}(\mathbf{k},s)= 2W_k\frac{M^*_1M^*_2}{E^*_1E^*_2}
            \frac{\Lambda^{(1)}_+(\mathbf{P}+\mathbf{k})Q(\mb{k},\mb{P})
            \Lambda^{(2)}_+(\mathbf{P}-\mathbf{k})}{W_0^2-W_k^2}.
            \end{equation}
 We take this propagator into BS equation~\eqref{BS-4D} in three-dimension space,
             \begin{equation}\label{BbS-Eq-Pre}
             \mathcal{T}(\mathbf{q}',\mathbf{q}|\mathbf{P})=
             \mathcal{V}(\mathbf{q}',\mathbf{q}|\mathbf{P})+
             \int \frac{d^3k}{(2\pi)^3}2W_k\frac{M^*_1M^*_2}{E^*_1E^*_2}
             \mathcal{V}(\mathbf{q}',\mathbf{k}|\mathbf{P})
             Q(\mb{k},\mb{P})
             \frac{\Lambda^{(1)}_+(\mathbf{P}+\mathbf{k})
                   \Lambda^{(2)}_+(\mathbf{P}-\mathbf{k})}
             {W_0^2-W_k^2}
              \mathcal{T}(\mathbf{k},\mathbf{q}'|\mathbf{P}).
             \end{equation}
With expressions of Dirac spinor, the invariance amplitude, $\mathcal{T}$ and $NN$ potential, $\mathcal{V}$ can be written as Lorentz scalars,
            \begin{equation}\label{V->scalar}
            \begin{aligned}
              \bar{V}(\mathbf{q'},\mathbf{q})&=
              \bar{u}_1(\mathbf{P}+\mathbf{q})\bar{u}_2(\mathbf{P}-\mathbf{q})
              \mathcal{V}(\mathbf{q'},\mathbf{q}|\mathbf{P})
              u_1(\mathbf{P}+\mathbf{q})u_2(\mathbf{P}-\mathbf{q}),
              \\
              \bar{T}(\mathbf{q'},\mathbf{q}|\mathbf{P})&=
              \bar{u}_1(\mathbf{P}+\mathbf{q})\bar{u}_2(\mathbf{P}-\mathbf{q})
              \mathcal{T}(\mathbf{q'},\mathbf{q}|\mathbf{P})
              u_1(\mathbf{P}+\mathbf{q})u_2(\mathbf{P}-\mathbf{q}).
            \end{aligned}
            \end{equation}
Thus the BbS equation in nuclear medium, Eq.~\eqref{BbS-Eq-Pre}, is rewritten as
             \begin{equation}\label{BbS-Eq}
             \bar{T}(\mathbf{q}',\mathbf{q}|\mathbf{P})=
             \bar{V}(\mathbf{q}',\mathbf{q})+
             \int \frac{d^3k}{(2\pi)^3}\frac{M^*_1M^*_2}{E^*_1E^*_2}
             \bar{V}(\mathbf{q}',\mathbf{k})
             \frac{2W_k}{W_0+W_k}
             \frac{Q(\mb{k},\mb{P})}{W_0-W_k}
             \bar{T}(\mathbf{k},\mathbf{q}|\mathbf{P}).
             \end{equation}
 The in-medium scattering amplitude and $NN$ potential can be redefined with the normalization condition of spinor~\eqref{Norm},
            \[
            G(\mathbf{q}',\mathbf{q}|\mathbf{P}) =
            \frac{M^*_1}{E^*_1}\bar{T}(\mathbf{q}',\mathbf{q}|\mathbf{P})
            \frac{M^*_2}{E^*_2}\quad\text{and}\quad
            V(\mathbf{q}',\mathbf{q}) =
            \frac{M^*_1}{E^*_1}\bar{V}(\mathbf{q}',\mathbf{q})
            \frac{M^*_2}{E^*_2}.
            \]{}
Finally the Eq.~\eqref{BbS-Eq-Pre} can be simplified as a more compact form, 
             \begin{equation}\label{BbS-Eq-Q}
            G(\mathbf{q}',\mathbf{q}|\mathbf{P})=
            V(\mathbf{q}',\mathbf{q})+
             \int \frac{d^3k}{(2\pi)^3}
            V(\mathbf{q}',\mathbf{k})
             \frac{2W_k}{W_0+W_k}
             \frac{Q(\mb{k},\mb{P})}{W_0-W_k}
            G(\mathbf{k},\mathbf{q}|\mathbf{P}).
             \end{equation}

\section{The detailed formulas for asymmetric nuclear matter} 
In BbS equation, three momenta, $\mathbf{q}',\mathbf{q}$, and $\mathbf{P}$ must be treated. When the asymmetric nuclear matter is considered, the integrals about these momenta become very complicated, especially for the Pauli operator.  In conventional calculations of RBHF model, the Pauli operator $Q_{\tau_1\tau_2}$ in the propagator is replaced by its average over solid angle with different cases~\cite{tong18}.           
            
            For the case $\tau_1\tau_2 = pp~\text{or}~nn$:
            \begin{itemize}
              \item (a)~~$ 0<P\leqslant k_F^\tau$:
              \begin{align} 
                Q_{\tau\tau}^\text{av} =\left\{\begin{array}{lr}
                0 & k<\Gamma ,\\
                \frac{k^2-\Gamma^2}{2Pk} &  \quad \Gamma\leqslant k<k_F^\tau+P,\\
                    1 & k\geqslant P+k_F^\tau, \\
                \end{array}\right.
              \end{align}
              \item{(b)}~~$P>k_F^\tau$:
              \[
                 Q_{\tau\tau}^\text{av} = 0.
              \]
            \end{itemize}
            with $\Gamma = \sqrt{k_F^{\tau 2} -P^2}$.

            \noindent
            For the case $\tau_1\tau_2=np~\text{or}~pn$:
            \begin{itemize}
                \item (a)~~$0\leqslant P \leqslant (k_F^n-k_F^p)/2$:
                \begin{align}
                Q_{np}^\text{av} =\left\{\begin{array}{lr}
                0 & k<k_F^n-P,\\
                \frac{(k+P)^2 -k_F^{n 2} }{4Pk}
                &\quad  k_F^n-P\leqslant k<k_F^n+P,\\
                    1 & k\geqslant P+k_F^n. \\
                \end{array}\right.
                \end{align}
                \item (b)~~$(k_F^n-k_F^p)/2P \leqslant (k_F^n+k_F^p)/2 $:
                \begin{align}
                Q_{np}^\text{av} =\left\{\begin{array}{lr}
                0 & k<\Gamma, \\
                \frac{k^2 -\Gamma^2}{2Pk}  &\Gamma\leqslant k <k_F^p+P,\\
                \frac{(k + P)^2-k_F^{n2}}{4Pk}
                & \quad k_F^p+P\leqslant k <k_F^n+P,\\
                1 & k\geqslant k_F^n+P.
                \end{array}\right.
                \end{align}
                \item (c) ~~ $P>(k_F^p+k_F^n)/2$:
                \[
                Q_{np}^\text{av} = 0,
                \]
            \end{itemize}
            with $\Gamma = \sqrt{(k_F^{p 2}+k_F^{n 2})/2 -P^2}$.
            
With this approximation, the solid angle dependence is removed in the integral of Eq.~\eqref{BbS-Gmatrix} at partial wave representation,
            \begin{equation} \label{BbS-Gmatrix-PW}
                G_{\tau_1\tau_2,\ell_1 \ell_2}^{\alpha_J}(q',q|P) =
                V^{\alpha_J}_{\tau_1\tau_2,\ell_1\ell_2}  (q',q)
                + \sum_{\ell'}\int k^2dk~V^{\alpha_j}_{\tau_1\tau_2,\ell_1\ell'}
                (q',k)\frac{2W_k}{W_0+W_k}\frac{Q^\text{av}_{\tau_1\tau_2}(k,P)}{W_0 - W_k}
                G^{\alpha_J}_{\tau_1\tau_2,\ell'\ell_2}(k,q|P),
            \end{equation}
where $\alpha_j$ indicate six possible $|LSJ\ket$ states with a fixed total angular momentum $J$. At the same time, the single-particle potential \eqref{SP-Pot} is transformed into center-of-mass frame, and then decomposed into partial wave $|LSJ\ket$ states. Its explicit expression is shown as~\cite{alonso03},
            \begin{align}\label{SP-Num}
                U_{\tau_1}(p)  = \sum_{\tau_2=p, n} \int_0^{\frac{k_F^{\tau_2}+p}{2}}
                dq \cdot  q^2C_{\tau_2}(p,q)\left[
                t^T_{\tau_1\tau_2}\sum_{j,\alpha_j}(2J+1)
                G^{\alpha_j}_{\tau_1\tau_2}(q_i|P^\text{av}_{\tau_2}(p,q))\right].
            \end{align}
 The coefficients $t_{\tau_1\tau_2}^T$ are concerned with different isospins,
            \[
                t^1_{nn} = t^1_{pp} = 1, \quad t^0_{pn} = t^0_{pn} = \frac{1}{2}, \quad
                \mathrm{otherwise:} ~~t_{\tau_1\tau_2}^T = 0.
            \]
The factor $C_\tau(p,q)$ are related with the Fermi momentum of nucleon,
            \begin{equation} \label{SP-Pot-C}
            C_\tau = \left\{\begin{array}{lr}
                            8  &
                            0<q\leqslant\frac{k^\tau_\text{F}-p}{2},\\
                            \frac{k_F^{\tau 2} -(p-2q)^2}{pq}&\quad
                            \frac{k^\tau_F-p}{2}<q\leqslant
                            \frac{k^\tau_F+p}{2}.\\
                        \end{array}\right.
            \end{equation}
The averaged total momentum in Pauli operator is given by
            \begin{equation}\label{SP-Pav1}
            \begin{aligned} 
            p\leqslant k_F :\quad
            P^\text{av}_\tau &= \left\{\begin{array}{lr}
                            \sqrt{p^2+q^2} & 0<q\leqslant\frac{k_F^\tau-p}{2},\\
                            \\
                            \frac{1}{2}\sqrt{3p^2+k^{\tau 2}_F-4pq}
                            &\quad \frac{k^\tau_F-p}{2}<q\leqslant\frac{k^\tau_F+p}{2}.\\
                        \end{array}\right. \\
            \end{aligned}
            \end{equation}
            \begin{equation}\label{SP-Pav2}
                        p> k_F :\quad
            P^\text{av}_\tau =\frac{1}{2}\sqrt{3p^2+k^{\tau 2}_F-4pq}
            \qquad \frac{p-k^\tau_F}{2}<q\leqslant \frac{k^\tau_F+p}{2}.
            \end{equation}
            otherwise $P^\text{av}_\tau=0$.

 When the total energy of nuclear matter is evaluated, the relevant integrals can also be performed in center-of-mass frame and taken as the partial-wave decomposition~\cite{tong18},
            \begin{equation} \label{Ev-Num}
            \begin{aligned}
                &\frac{1}{2n_b}
                \sum_{\tau_1\tau_2} \sum_{ss'}
                \int^{p\leqslant k^{\tau_1}_F} \frac{d^3p}{(2\pi)^3}
                \int^{p'\leqslant k^{\tau_2}_F} \frac{d^3p'}{(2\pi)^3}
                \langle ps,p's'|G_{\tau_1\tau_2}(W_{\tau_1\tau_2})
                |ps,p's'\rangle
                \\
                =&\frac{1}{(2\pi)^3}\frac{8}{2n_b}
                \sum_{\tau_1\tau_2}\int^{(k^{\tau_1}_F
                +k^{\tau_2}_\text{F})/2} d^3q \int^{{\substack{
                |\mathbf{P}+\mathbf{q}|\leqslant k_F^{\tau_1} \\
                |\mathbf{P}-\mathbf{q}|\leqslant k_F^{\tau_2}} }}
                d^3P~ t^T_{\tau_1\tau_2}\sum_{J,\alpha_j}(2J+1)
                G_{\tau_1\tau_2}^{\alpha_j}(q,q|P).
            \end{aligned}
            \end{equation}
The solid angle integration for center-of-mass momenta in last line of this equation is divided into following cases.
            
            For the case of $\tau_1\tau_2=pp,~nn$,
            \begin{align}\label{Ev-D1}
            \int d\Omega_P=\left\{\begin{array}{lr}
                     2 &
                     0\leqslant P\leqslant k_\text{F}^\tau-q, \\
                     \frac{\Gamma^2-P^2}{Pq} & \quad
                     k_F^\tau-q<P\leqslant \Gamma,  \\
                     0  & P>\Gamma,\\
                     \end{array}\right.
            \end{align}
            with $\Gamma = \sqrt{k_F^{\tau 2}-q^2}$.
            
            For the case of $\tau_1\tau_2= np,~pn$,
            \begin{align}\label{Ev-D2}
            \int d\Omega_P = \left\{\begin{array}{lr}
                    2 & 0\leqslant P\leqslant k_F^p-q,\\
                    \frac{k_F^{p2}-(q-P)^2}{2Pq}  & \quad
                     k_F^p-q<P\leqslant k_F^n-q,\\
                     \frac{\Gamma^2-P^2}{Pq} &
                     k_F^n-q<P\leqslant \Gamma,  \\
                     0  & P>\Gamma,\\
                     \end{array}\right.
            \end{align}
            with $\Gamma =\sqrt{(k_\text{F}^{p2}+k_\text{F}^{n2})/2-q^2}$.


\begin{thebibliography}{99}
	\bibitem{baldo99} M. Baldo (Ed.), Nuclear Methods and Nuclear Equation of State in International Review of
	Nuclear Physics, Vol. \textbf{8} (World Scientific Publishing Company, 1999).
	
	\bibitem{klupfel09} P. Kl\"upfel, P.-G. Reinhard, T. J. Burvenich,  and J. A. Maruhn,  Phys. Rev. C \textbf{79}, 034310 (2009).
	
	\bibitem{rocamaza18} X. Roca-Maza and N. Paar, Prog. Part. Nucl. Phys. \textbf{101}, 96 (2018).
	
	\bibitem{oertel17} M. Oertel, M. Hempel, T. Kl\"{a}hn, and S. Typel. Rev. Mod. Phys. {\bf 89}, 015007 (2017). 
	
	\bibitem{abbott17a} B. P. Abbott {\it et al.}, (LIGO Scientific and Virgo Collaboration), Phys. Rev. Lett. {\bf 119}, 161101 (2017).
	
	\bibitem{abbott17b} B. P. Abbott {\it et al.}, (Virgo, Fermi-GBM, INTEGRAL, and LIGO Scientific Collaboration), Astrophys. J. {\bf 848}, L13 (2017).
	
	\bibitem{abbott17c} B. P. Abbott {\it et al.}, Astrophys. J. {\bf 848}, L12 (2017).
	
	\bibitem{goldstein17} A. Goldstein {\it et al.}, Astrophys. J. {\bf 848}, L14 (2017).
	
	\bibitem{abbott18a} B. P. Abbott {\it et al.}, (LIGO Scientific and Virgo Collaboration), Phys. Rev. Lett. {\bf 121}, 161101 (2018).
	
	\bibitem{li08} B.A. Li, L.W. Chen, and C.M. Ko, Phys. Rep. {\bf 464}, 113 (2008).
	
	\bibitem{li19} B. A. Li, P. G. Krastev, D. H. Wen, and N. B. Zhang,  Eur. Phys. J. A {\bf 55}, 117 (2019).
	
	\bibitem{shen98a} H. Shen, H. Toki, K. Oyamatsu, and K. Sumiyoshi, Nucl. Phys. A {\bf 637}, 435 (1998). 
	
	\bibitem{shen98b} H. Shen, H. Toki, K. Oyamatsu, and K. Sumiyoshi, Prog. Theor. Phys. {\bf 100}, 1013 (1998).
	
	\bibitem{shen02} H. Shen, Phys. Rev. C {\bf 65}, 035802 (2002).
	
	\bibitem{shen11} H. Shen, H. Toki, K. Oyamatsu and K Sumiyoshi, Astrophys. J. Suppl. {\bf 197}, 20 (2011).
	
	\bibitem{euler37} H. Euler, Z. Phys. {\bf 105}, 553 (1937).
	
	\bibitem{jastrow51} R. Jastrow, Phys. Rev. {\bf 81}, 165 (1951).
	
	\bibitem{brueckner54} K. A. Brueckner, C. A. Levinson, and H. M. Mahmound, Phys. Rev. {\bf 95}, 217 (1954).
	
	\bibitem{bethe56} H. A. Bethe, Phys. Rev. {\bf 103}, 1353 (1956).
	
	\bibitem{jastrow55} R. Jastrow, Phys. Rev. {\bf 98}, 1479 (1955).
	
	\bibitem{li06} Z. H. Li, U. Lombardo, H.-J. Schulze, W. Zuo, L. W. Chen, and H. R. Ma,  Phys. Rev. C {\bf 74}, 047304 (2006).
	
	\bibitem{baldo07} M. Baldo, C. Maieron, J. Phys. G {\bf 34}, R243 (2007).
	
	\bibitem{baldo16} M. Baldo and G. F. Burgio, Prog. Part. Nucl. Phys. {\bf 91}, 203 (2016).
	
	\bibitem{akmal98} A. Akmal, V. R. Pandharipande, and D. G. Ravenhall,  Phys. Rev. C {\bf 58}, 1804 (1998).
	
	\bibitem{carlson15}J. Carlson, S. Gandolfi, F. Pederiva, Steven C. Pieper, R. Schiavilla, K. E. Schmidt, and R. B. Wiringa, Rev. Mod. Phys. {\bf 87}, 1067 (2015).
	
	\bibitem{dickhoff04}W. H. Dickhoff and C. Barbieri, Prog. Part. Nucl. Phys. {\bf 52}, 377 (2004).
	
	\bibitem{hagen14a} G. Hagen, T. Papenbrock, M. Hjorth-Jensen, and D. J. Dean, Rep. Prog. Phys. {\bf 77}, 096302 (2014).
	
	\bibitem{hagen14} G. Hagen, T. Papenbrock, A. Ekstr\"{o}m, K. A. Wendt, G. Baardsen, S. Gandolfi, M. Hjorth-Jensen, and C. J. Horowitz, Phys. Rev. C {\bf 89}, 014319 (2014).
	
	\bibitem{carbone13} A. Carbone, A. Polls, and A. Rios, Phys. Rev. C {\bf 88}, 044302 (2013).
	
	\bibitem{carbone14} A. Carbone, A. Rios, and A. Polls, Phys. Rev. C {\bf 90}, 054322 (2014).
	
	\bibitem{drischler14} C. Drischler, V. Som$\grave{a}$, and A. Schwenk, Phys. Rev. C {\bf 89}, 025806 (2014).
	
	\bibitem{drews15} M. Drews and W. Weise, Phys. Rev. C {\bf 91}, 035802 (2015).
	
	\bibitem{drews16} M. Drews and W. Weise, Prog. Part. Nucl. Phys. {\bf 93}, 69 (2017).
	
	\bibitem{modarres93} M. Modarres, J. Phys. G: Nucl. Part. Phys. {\bf 19}, 1349 (1993).
	
	\bibitem{stoks94}V. G. J. Stoks, R. A. M. Klomp, C. P. F. Terheggen, and J. J. de Swart, Phys. Rev. C {\bf 49}, 2950 (1994).
	
	\bibitem{wiringa95} R. B. Wiringa, V. G. J. Stoks, and R. Schiavilla, Phys. Rev. C {\bf 51}, 38 (1995).
	
	\bibitem{machleidt01} R. Machleidt, Phys. Rev. C {\bf 63}, 024001 (2001).
	
	\bibitem{entem03} D. R. Entem and R. Machleidt, Phys. Rev. C {\bf 68}, 041001(R) (2003).
	
	\bibitem{epelbaum05} E. Epelbaum, W. Gl\"{o}ckle, and U.-G. Mei\ss ner, Nucl. Phys. A {\bf 747}, 362 (2005).
	
	\bibitem{epelbaum15a} E. Epelbaum, H. Krebs, and U.-G. Mei\ss ner, Eur. Phys. J. A {\bf 51}, 53 (2015).
	
	\bibitem{epelbaum15b} E. Epelbaum, H. Krebs, and U.-G. Mei\ss ner, Phys. Rev. Lett. {\bf 115}, 122301 (2015).
	
	\bibitem{entem15} D. R. Entem, N. Kaiser, R. Machleidt, and Y. Nosyk, Phys. Rev. C {\bf 91}, 014002 (2015).
	
	\bibitem{entem17} D. R. Entem, R. Machleidt, and Y. Nosyk, Phys. Rev. C {\bf 96}, 024004 (2017).
	
	\bibitem{reinert18} P. Reinert, H. Krebs, and E. Epelbaum, Eur. Phys. J. A {\bf 54}, 86 (2018).
	
	\bibitem{hu17} J. Hu, Y. Zhang, E. Epelbaum, Ulf-G. Meissner, and J. Meng, Phys. Rev. C {\bf 96}, 034307(2017).
	
	\bibitem{sammarruca18} F. Sammarruca, L. E. Marcucci, L. Coraggio, J. W. Holt, N. Itaco, R. Machleidt, arXiv:1807.06640
	
	\bibitem{logoteta19} D. Logoteta, Phys. Rev. C {\bf 100}, 045803 (2019).
	
	\bibitem{ansatasio83} M. R. Ansatasio, L. S. Celenza, W. S. Pong, and C. M. Shakin, Phys. Rep. {\bf 100}, 327 (1983).
	
	\bibitem{horowitz87} C. J. Horowitz and B. D. Serot, Nucl. Phys. A {\bf464}, 613 (1987).
	
	\bibitem{brockmann90} R. Brockmann and R. Machleidt,  Phys. Rev. C {\bf 42}, 1965 (1990).
	
	\bibitem{lizh08} Z. H. Li, U. Lombardo, H.-J. Schulze, and W. Zuo,  Phys. Rev. C {\bf 77}, 034316 (2008).
	
	\bibitem{alonso03} D. Alonso and F. Sammarruca, Phys. Rev. C {\bf 68}, 054305 (2003).
	
	\bibitem{krastev06} P. G. Krastev and F. Sammarruca,  Phys. Rev. C {\bf 74}, 025808 (2006).
	
	\bibitem{sammarruca10} F. Sammarruca, Int. J. Mod. Phys. E {\bf 19}, 1259 (2010).
	
	\bibitem{dalen10} E. N. E. Dalen and H. Muether, Int. J. Mod. Phys. E {\bf 19}, 2077 (2010).
	
	\bibitem{shen16} S. Shen, J. Hu, H. Liang, J. Meng, P. Ring, and S. Zhang, Chin. Phys. Lett. {\bf 33},  102103 (2016).
	
	\bibitem{shen17}  S. Shen, H. Liang, J. Meng,  P. Ring, and S. Q. Zhang, Phys. Rev. C {\bf 96}, 014316(2017).
	
	\bibitem{shen19} S. Shen, H. Liang, W. Long, J. Meng, and P. Ring, Prog. Part. Nucl. Phys. {\bf 109}, 103713 (2019).
	
	\bibitem{tong18} H. Tong, X. L. Ren, P. Ring, S. H. Shen, S. B. Wang, and J. Meng, Phys. Rev. C {\bf98}, 054302 (2018).
	
	\bibitem{machleidt87} R. Machleidt, K. Holinde, and Ch. Elster, Phys. Rep. {\bf 149}, 1 (1987).
	
	\bibitem{machleidt89} R. Machleidt, Adv. Nucl. Phys.  {\bf 19}, 189 (1989). 
	
	\bibitem{hu19} J. Hu, P. We, and, Y. Zhang, Phys. Lett. B {\bf 798}, 134982 (2019).
	
	\bibitem{wang19} C. Wang, J. Hu, Y. Zhang, and H. Shen, Chin, Phys. C {\bf 43}, 114107 (2019).
	
	\bibitem{sammarruca12} F. Sammarruca, L. White, and B. Chen, Eur. Phys. J. A {\bf 48}, 181 (2012).
	
	\bibitem{dalen10a} E. N. E. Dalen and H. M\"uther, Phys. Rev. C {\bf82}, 014319  (2010). 
	
	\bibitem{salpeter51} E. E. Salpeter and H. A. Bethe, Phys. Rev. {\bf 84}, 1232 (1951).
	
	\bibitem{blankenbecler66} R. Blankenbecler and R. Sugar, Phys. Rev. {\bf 142}, 1051 (1966).
	
	\bibitem{thompson70} R. H. Thompson, Phys. Rev. D {\bf 1}, 110 (1970).
	
	\bibitem{kadyshevsky68} V. G. Kadyshevsky, Nucl. Phys. B {\bf 6}, 125 (1968).
	
	\bibitem{deltuva03} A. Deltuva, R. Machleidt, and P. U. Sauer, Phys. Rev. C {\bf 68}, 024005 (2003).
	
	\bibitem{logoteta16} D. Logoteta, I. Bombaci, and A. Kievsky, Phys. Rev. C {\bf 94}, 064001 (2016). 
	
	\bibitem{ekstrom18} A. Ekstr\"om, G. Hagen, T. D. Morris, T. Papenbrock, and P. D. Schwartz, Phys. Rev. C {\bf 97}, 024332 
	
	\bibitem{hu10} J. Hu, H. Toki, W. Wen, and H. Shen,  Phys. Lett. B {\bf 687}, 271 (2010).
	
	\bibitem{wang12} Y. Wang, J. Hu, H. Toki, and H. Shen, Prog. Theo. Phys. {\bf 127}, 739 (2012).
	
	\bibitem{zhang19} Y. Zhang, P. Liu, and J. Hu, arXiv:1910.11765.
	
	\bibitem{coester70} F. Coester, S. Cohen, B. Day, and C. M. Vincent, Phys. Rev. C {\bf 1}, 769 (1970).
	
	\bibitem{hu13} J. Hu, H. Toki, and Y. Ogawa, Prog. Theor. Exp. Phys. {\bf 103D02}, (2013).
		
	\bibitem{serot86} B. D. Serot and J. D. Walecka,  Adv. Nuc. Phys. {\bf 16}, 1 (1986).
	
	\bibitem{fuchs98} C. Fuchs, T. Waindzoch, A. Faessler, and D. S. Kosov, Phys. Rev. C {\bf 58}, 2022 (1998).
	
	\bibitem{bouyssy87} A. Bouyssy, J. -F Mathiot, N. Van Giai, and S. Marcos, Phys. Rev. C {\bf 36}, 380 (1987).
	
	\bibitem{long06}W. H. Long, N. Van Giai, and J. Meng, Phys. Lett. B {\bf 640}, 150 (2006).
	
	\bibitem{long07}W. H. Long, H. Sagawa, N. V. Giai, and J. Meng, Phys. Rev. C {\bf 76}, 034314 (2007).
	
	\bibitem{gandolfi12} S. Gandolfi, J. Carlson, and S. Reddy, Phys. Rev. C {\bf 85}, 032801(R) (2012).
	
	\bibitem{danielewicz09} P. Danielewicz and J. Lee, Nucl. Phys. A {\bf 818} 36 (2009).
	
	\bibitem{garg18} U. Garg and G. Col\'o, Prog. Part. Nucl. Phys. {\bf 101}, 55 (2018).
	
	\bibitem{russotto16} P. Russotto {\it et al.}, Phys. Rev. C {\bf 94}, 034608 (2016).
	
	\bibitem{holt17} J. W. Holt and N. Kaiser, Phys. Rev. C {\bf 95}, 034326 (2017).
	
	\bibitem{lim19} Y. Lim and J. W. Holt, Eur. Phys. J. A {\bf 55}, 209 (2019).
	
	\bibitem{tong19}H. Tong, P. W. Zhao, and J. Meng, arXiv:1903.05938.
	
	
\end{thebibliography}
\end{document}